\documentclass[12pt,preprint]{aastex}
\shorttitle{Galaxies and Damped Ly$\alpha$ Absorbers at $z\sim3$}
\shortauthors{Cooke et al.}

\begin{document}

\title{Survey for Galaxies Associated with $z\sim3$ Damped Lyman $\alpha$ 
Systems I: Spectroscopic Calibration of $u'$BVRI Photometric Selection}

\author{Jeff Cooke and Arthur M. Wolfe \altaffilmark{1} \altaffilmark{2}}
\email{cooke@physics.ucsd.edu}
\email{awolfe@ucsd.edu}

\author{Jason X. Prochaska \altaffilmark{1} \altaffilmark{3}}
\email{xavier@ucolick.edu}

\and
 
\author{Eric Gawiser \altaffilmark{1} \altaffilmark{4} \altaffilmark{5}}
\email{gawiser@astro.yale.edu}

\altaffiltext{1}{Visiting Astronomer, W. M. Keck Telescope. The Keck
  Observatory is a joint facility of the University of California, the
  California Institute of Technology, and NASA and was made possible by 
  the generous financial support of the W. M. Keck Foundation.}
\altaffiltext{2}{Department of Physics and Center for Astrophysics and
  Space Sciences, University of California, San Diego, 9500 Gilman
  Drive, Code 0424, La Jolla, CA 92093-0424.}
\altaffiltext{3}{University of California Observatories, Lick
  Observatory, Natural Sciences II Annex, UC Santa Cruz, Santa Cruz CA
  95064.}
\altaffiltext{4}{NSF Postdoctoral Fellow, Yale University, P.O. Box
  208101, New Haven, CT 06520.}
\altaffiltext{5}{Andes Prize Fellow, Universidad de Chile, Casilla
  36-D, Santiago, Chile.}

%---------------------------------------------------------------------

\begin{abstract}
We present a survey for $z\sim3$ Lyman break galaxies (LBGs)
associated with damped Lyman $\alpha$ systems (DLAs) with the primary 
purpose of determining the DLA-LBG cross-correlation.  This paper 
describes the acquisition and analysis of imaging and spectroscopic 
data of 9 quasar fields having 11 known $z\sim3$ DLAs covering an 
area of 465 arcmin$^2$.  Using deep $u'$BVRI images, 
796 LBG candidates to an apparent R$_{AB}$ magnitude of 25.5 were 
photometrically selected from 17,343 sources detected in the field.  
Spectroscopic observations of 529 LBG candidates using Keck $LRIS$ 
yielded 339 redshifts.  We have conservatively identified 211 $z > 2$ 
objects with $\langle z\rangle=3.02\pm0.32$.  We discuss our method 
of $z\sim3$ LBG identification and present a model of the $u'$BVRI 
photometric selection function.  We use the 339 spectra to 
evaluate our $u'$BVRI $z\sim3$ Lyman break photometric selection 
technique.
\end{abstract}

% --- intergalactic medium

%---------------------------------------------------------------------------
%   TEXT
%---------------------------------------------------------------------------

\section{INTRODUCTION}
\vspace{0.05in}

Over the last decade, high redshift surveys to detect Lyman break
galaxies (LBGs) by photometric pre-selection, \citep {s96,st96,l97}, 
have been 
successful in compiling catalogs of $\sim1000$ galaxy spectra at 
$z\sim3$.  This has become an efficient technique to obtain
statistical information of high redshift galaxies via their emission 
properties.  Analysis of the LBG spatial distribution shows
significant clustering which supports the belief that they are 
biased tracers of mass and form in overdense regions. The LBG 
auto-correlation function \citep {s98,a98,a03} and best-fit star
formation models \citep {aes01} suggest LBGs are massive objects that 
most likely evolve into early-type galaxies by the present day.  LBG
surveys, such as the spectroscopic sample, are magnitude limited and
only probe the higher end of the proto-galactic mass distribution. 

By contrast, quasar (QSO) absorption line systems represent a
population of high redshift protogalaxies whose detections are not 
biased by their intrinsic magnitude.  Their detection is dependent 
only on the strengths of their absorption line features and
the brightness of the background QSO.  The Damped Lyman $\alpha$
systems (DLAs) are the highest column density absorbers, defined to
have N(HI) $\ge 2\times10^{20}$ atoms cm$^{-2}$ as established by 
\citet {w86}.  DLAs are drawn from the bulk of the proto-galactic mass 
distribution and harbor the majority of the neutral hydrogen present
in the universe.  The agreement between the comoving mass 
density of neutral gas at $z > 2$ and the mass density of visible
stars in local galactic disks \citep {w95} lends credibility to the
idea that DLAs evolve to form present-day galaxies such as the Milky
Way \citep {k96}.  However, this agreement in baryon mass cannot 
differentiate between passive models in which DLAs are protogalaxies
having properties similar to those of local galaxies with constant 
comoving densities from CDM models in which DLAs are numerous,
low-mass objects with comoving densities that decrease with time.

These two interpretations have been substantiated largely from 
detailed analyses of DLA gas kinematics. \citet {pw97,pw98} and \citet
{w00a,w00b} best described the asymmetries in the absorption line
profiles using large thick centrifugally-supported disks drawn from a 
Schecter luminosity/mass function that obey the Tully-Fisher relation.  
Conversely, hierarchical CDM scenarios confine DLAs to compact
low-mass sub-L* systems. Models of irregular merging clumps are shown to
reproduce the observed absorption line kinematics as well \citep {h98,h00}.  
A reliable determination of the mass of typical individual DLAs is
needed to break the degeneracy between these interpretations and 
provide a vital parameter for numerical models of galaxy formation.

DLAs offer a wealth of information from their absorption line 
characteristics but their baryonic mass cannot be determined without
knowing their typical cross-sectional area.  The incidence of DLAs
along quasar lines-of-sight only determines the product of their 
comoving densities and cross-sections.  It is possible, however, to 
infer the dark-matter mass of typical DLAs in the context of CDM 
cosmology.  
Hierarchical structure formation models predict that the most massive 
galaxies will preferentially form clustered together in regions of
mass overdensity.  Under the assumption of a Gaussian distribution of
density fluctuations in the density field, nearly all of the density 
fluctuations that give rise to massive galaxies (with the exception of 
those on the high-end tail of the distribution) have initial density 
contrasts below that necessary for collapse against the Hubble flow.  
However, these density fluctuations receive a boost in density when
superimposed onto a larger wavelength fluctuation from an over-dense 
region.  In this situation, the number of the fluctuations exceeding 
the threshold of collapse that correspond to massive galaxies  
is increased with the tendency of these fluctuations to be 
clustered near the peak of the underlying overdense region.
Conversely, density fluctuations that give rise to low-mass galaxies
have a relatively high initial density contrast and are predicted to 
form more uniformly throughout space.  

An enhancement in the clustering of galaxies is referred to as bias 
\citep {k84}.  Therefore, the measurement of the bias of a population
of galaxies provides a means to infer their mass distribution.
The bias of a particular sample of galaxies can be derived from their
spatial auto-correlation or by their cross-correlation with another
population of galaxies with known bias.   DLAs are detected in
absorption in quasar lines-of-sight and their sparse sampling at 
$z\sim3$ makes the measurement of the DLA auto-correlation function 
difficult. Since spectral analysis from previous surveys for LBGs in 
emission has yielded the LBG auto-correlation function and established 
the LBG galaxy bias at $z\sim3$ \citep{a03}, an effective way, and
perhaps the only way, to determine the DLA bias is through the DLA-LBG 
cross-correlation.  

In general, the Poisson probability that two volume elements, $dV_1$
and $dV_2$, in the field separated by a distance $r$ are occupied by a 
DLA and an LBG is:
\begin {equation}
dP(r) = n_{DLA} n_{LBG}~dV_{DLA} dV_{LBG},
\end {equation}
where $n_{DLA}$ and $n_{LBG}$ are the mean number densities of each 
population.  The presence of a DLA in one volume element enhances the 
probability of the presence of an LBG in another volume element by an
amount described by the DLA-LBG cross-correlation function
$\xi_{DLA-LBG}(r)$. 
%  the probability that two volumes are occupied by an LBG is expressed as:
%  \begin {equation}
%  dP(r) = n_{LBG}^2~[ 1 + \xi_{LBG-LBG}(r)]~dV_1 dV_2.
%  \end {equation} 
%  Similar to the above expression, 
Therefore, the probability of finding an LBG in a volume element at a 
distance $r$ from a known DLA is expressed as:
\begin {equation}
dP(r) = n_{DLA} n_{LBG}~[ 1 + \xi_{DLA-LBG}(r)]~dV_{DLA} dV_{LBG}.
\end {equation}
%  Cosmological models, such as $\Lambda$CDM, determine the dark matter
%  correlation function $\xi_{mass}$ \citep{pea97}.  The LBG
%  auto-correlation function represents the extent in which luminous
%  galaxies trace the underlying distribution of mass and is therefore a 
%  biased indicator of the mass correlation function.  If the biasing is 
%  due to the fact that LBGs form in overdense regions, the LBG
%  auto-correlation function can be related to the mass correlation
%  function by the approximately linear relation:
%  \begin {equation}
%  \xi_{LBG-LBG} = b_{LBG}^2~ \xi_{CDM} 
%  \end {equation}
%  where $b_{LBG}$ is the bias parameter for LBGs and has been measured
%  \citep{a03}.  The cross-correlation between DLAs and LBGs then becomes:
At high redshift, where a linear bias to the underlying dark matter 
correlation function $\xi_{DM}$ \citep{pea97} is a reasonable model,
the DLA-LBG cross-correlation is:
\begin {equation}
\xi_{DLA-LBG} = b_{DLA}~b_{LBG}~ \xi_{DM} ,
\end {equation}
where $b_{DLA}$ and $b_{LBG}$ are the bias parameters for the two
populations.  Since the LBG bias at $z\sim3$ is known, and once the
DLA-LBG cross-correlation is measured, the dark matter mass of 
typical individual DLAs can be inferred directly from the resulting
value of the DLA bias using the approach of \citet{m98}.  
%This method 
%to probe the mass of high redshift DLAs will be discussed in greater 
%detail in Paper II.  

To measure the DLA-LBG cross-correlation function, we use a
photometric selection technique similar to previous successful
surveys, \citet {s98}, to efficiently detect LBGs in QSO fields with 
known DLAs.  We are able to construct the DLA-LBG spatial 
cross-correlation function from both angular and spectroscopic 
measurements.  Recently, \citet{bl03} estimated the two-dimensional 
DLA-LBG angular cross-correlation function in three wide fields using 
photometric redshifts.  They found a positive DLA-LBG clustering 
amplitude to a $>95\%$ confidence level based on Monte Carlo 
simulations.  Their photometric redshift measurements constrained 
their results to spatial clustering within redshift slices on the 
order of their photometric redshift uncertainties $(\Delta z\sim0.2)$.  
Conversely, \citet{a03} measured the mean overdensity of LBGs within 
cylindrical cells with a height $\Delta z=0.025$ and radius of up to 
$\Delta\theta=265\arcsec$ centered on the four DLAs in their 
spectroscopic sample at $z\sim3$.  They compared this mean overdensity 
with the mean overdensity of LBGs within similar cylindrical cells 
centered on LBGs drawn from their complete sample that had similar 
redshifts as two of their DLAs.  From their analysis, they argued that the 
three-dimensional DLA-LBG cross-correlation is weak and concluded that 
DLAs and LBGs do not have similar spatial distributions.  Since the 
statistical accuracy of both investigations was limited, neither confirms, 
nor rules out a significant value of the DLA bias.

Because DLAs may be a population of protogalaxies having a range of 
masses, a statistical determination of the DLA-LBG cross-correlation 
function from multiple DLA fields is required.  This 
paper presents the results of $u'$BVRI photometric selection of LBG 
candidates in a survey of 9 QSO fields containing 11 known $z\sim3$
DLAs.  We address the efficiency of using $u'$BVRI filters to detect 
LBGs and apply 339 spectra to calibrate the $u'$BVRI photometric 
selection process.  The $u'$BVRI filters are naturally suited to
select Lyman break objects at $z\sim3$ and are distinct from the $U_n G 
{\cal R}$ filter set of \citet {s92} in that they have the advantage
of being readily available at most observatories.  
%Although the  $U_n G 
%{\cal R}$ filters of \citet{s92} have been shown to be very effective 
%in selecting a large number of $z\sim3$ Lyman break objects, they were 
%originally designed for a system at $z=3.4$ and are not as generally 
%accessible.  

This marks the second step in an observing program designed to measure 
the DLA bias by searching for galaxies in emission that are clustered
with DLAs or directly responsible for the damped absorption.  Previous 
work by \citet {g01} and \citet {p02} detail the acquisition and
analysis of galaxies associated with DLAs at $z\sim4$ using deep BRI 
imaging.  The $z\sim4$ regime is more challenging due to a relatively
low LBG number density to R$_{AB}<25$, a large number of 
low-$z$ interlopers having similar colors, and the fact that the 
relevant interstellar absorption features are redshifted to
wavelengths that coincide with a dense array of bright night-sky
emission lines.  By contrast, at $z\sim3$ there are three times as
many LBGs at a given magnitude, fewer low-$z$ objects with similar
colors, and the expected Lyman $\alpha$ and metal absorption line
features are redshifted to a region with few night-sky emission lines 
and high CCD quantum efficiency.  

This paper is organized as follows: \S~2 describes the deep imaging 
observations and reduction process, \S~3 discusses the photometry and 
the photometric color selection, \S~4 describes the spectroscopic 
observations and analysis, \S~5 addresses the use of spectroscopically 
identified galaxies to refine the selection criteria, and \S~6
provides a brief summary.  The DLA-LBG cross-correlation and
analysis are presented in Paper II.  All magnitudes
reported in this paper are given using the $AB_{95}$ \citep 
{f96} scale unless otherwise stated. We adopted the following 
transformations from the standard Johnson-Cousins magnitude system 
to the $AB_{95}$ system: $u'_{AB}$ = U$_{JC}$ + 0.70, B$_{AB}$ = B$_{JC}$ 
-- 0.15, V$_{AB}$ =  V$_{JC}$ -- 0.01, R$_{AB}$ = R$_{JC}$ + 0.18, 
and I$_{AB}$ =  I$_{JC}$ + 0.43. 

%---------------------------------------------------------------------

\section{IMAGING OBSERVATIONS}

We chose QSO fields that contained at least one DLA in the redshift 
path consistent with the LBG detection volumes probed by $u'$BVRI 
photometric color selection.  We also maintained a preference for 
systems near $z=3.0$ where the photometric selection is most efficient
and avoided systems with less than 3000 km sec$^{-1}$ separation from 
the background QSO.  Fields were chosen by 
their coordinate accessibility and, when more than one field was 
observable, we chose the field that would be observed through the
lowest airmass and in the direction of lowest Galactic reddening.  
These considerations minimized the attenuation of the light from the 
LBGs and eased their detection using reasonable integration times, 
especially in the $u'$-band.  Prior to this survey, there were 10
known $z\sim3$ DLAs in the selected 9 QSO sight lines.  Our spectroscopy
revealed a second DLA in the field PSS0808+5215 at $z$=2.936 and was 
confirmed using the moderate resolution Keck ESI data of \citet {p03}. 
Table~\ref{coords} lists the 9 QSO fields and 11 $z\sim3$ DLAs used 
in this survey.  Our $u'$BVRI color selection technique does not 
efficiently detect LBGs near the 4.18 DLA near PSS0957+3308.  Although the
imaging depths of the PSS0957+3308 field allowed for reasonable color 
selection of $z\sim4$ LBG candidates using BVRI color criteria similar 
to those described in \citet {g01}, no follow-up spectroscopy has been 
acquired to date for these candidates.

The imaging data were acquired during the interval 2000 April through
2003 November at the Palomar and Keck Observatories.  We used the 
Carnegie Observatories Spectrograph and Multi-Object Imaging Camera 
\citep {k98} on the $200''$ Hale telescope at Palomar to obtain deep 
images for the first two fields in our survey. Both the PSS1432+3940 
and JVAS2344+3433 fields were imaged in $U_n$BR filters using 
$COSMIC$ equipped with a Tektronix 2048$\times$2048 thinned CCD
($0.\arcsec286$ pixel$^{-1}$).  Subsequent observations in V
and I were obtained using the Low Resolution Imager and Spectrometer
\citep {o95}.  The $COSMIC$ field of view (FOV) is $\sim
9.7\times 9.7$ arcmin whereas $LRIS$ FOV is $\sim5\times 7$ arcmin.  
Two pointings with $LRIS$ provided $\sim$ 70\% coverage of each 
$COSMIC$ field. Table~\ref{imaging} presents a journal of the 
observations for the 9 fields in the survey.

All other fields in our survey were imaged wholly using $LRIS$ with
$u',(U_n)$,B,V,R,(R$_s$), and I filters.  $LRIS$ is a dual arm 
instrument with both red and blue sensitive sides. Prior to January, 
2001, all images were taken using the red arm equipped with a 
back-side illuminated Tektronix 2048$\times$2048 thinned CCD 
($0.\arcsec215$ pixel$^{-1}$).  From January, 2001 through June, 
2002, $U_n$ and B-band images were obtained using the blue arm of 
$LRIS$ with a SITe 2048$\times$2048 thinned CCD ($0.\arcsec211$ 
pixel$^{-1}$). After June, 2002, $u'$ and B-band data were obtained 
using $LRIS$ equipped with 2 Marconi 2048$\times$4096 blue-sensitive 
CCDs ($0.\arcsec135$ pixel$^{-1}$).  

\begin{figure}
\begin{center}
\scalebox{0.65}[0.6]{\rotatebox{90}{\includegraphics{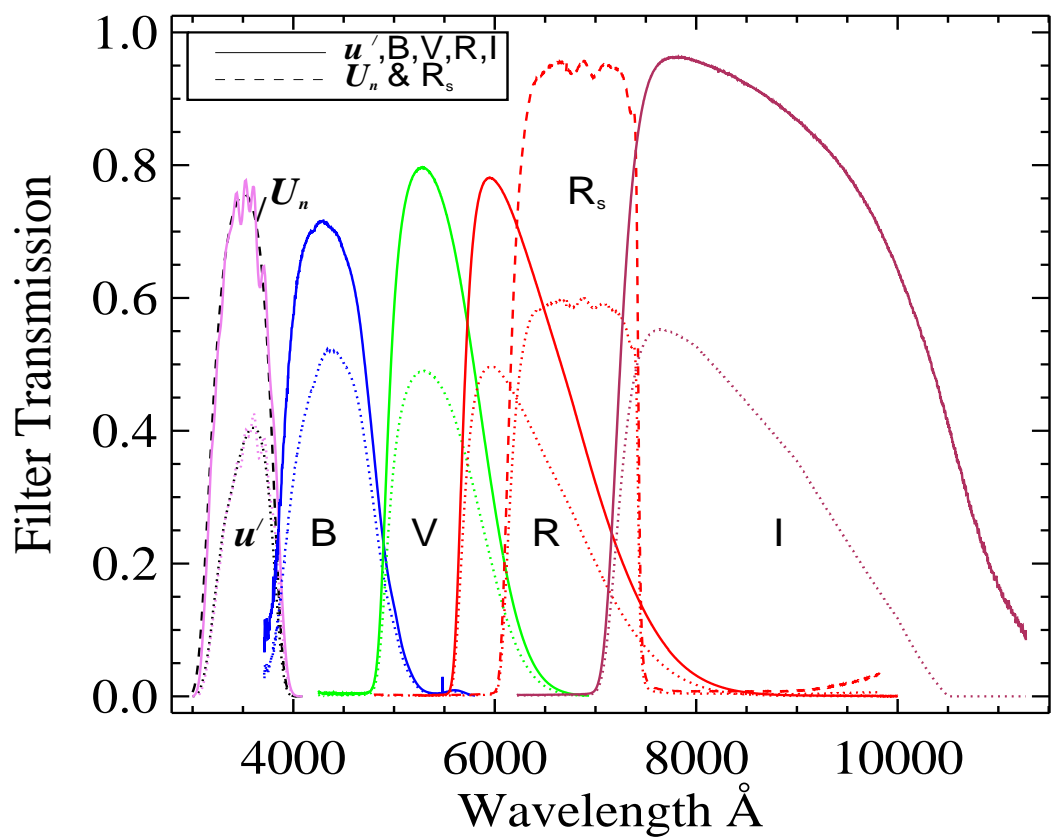}}}
\caption{Transmissions of the filters used in this survey. The
dotted curves indicate the effective transmissions corrected for 
atmosphere attenuation and the appropriate CCD quantum efficiencies.
The $U_n$ and R$_s$ filters were used in certain fields and are
denoted by the dashed curves. 
\label{5filters}} 
\end{center}
\end{figure}

Figure~\ref{5filters} plots the transmissions of the filters used in
this survey. The transmissions of the $U_n$ and R$_s$ filters are 
indicated by the dashed curves.  The $u'$ and $U_n$ filters were both 
used during the course of this survey as a result of their
availability at Keck.  Since the $u'$ and $U_n$ filters have nearly 
identical bandpasses, effective wavelengths, transmission
efficiencies, and application in this paper, the $U_n$ filter will be 
referred hereafter as the $u'$ filter for simplicity, unless otherwise 
noted.  The R$_s$ filter was used wholly or partially to image three 
fields as the survey progressed in an attempt to increase the number
of Lyman break candidate detections since it has a higher throughput 
and sharper transmission cutoffs than the R$_{JC}$. 
% Figure~\ref{allUs} 
%plots the $u'$ and $U_n$ filter transmissions for direct comparison.  

All observations were taken as close to the meridian as possible in an 
effort to minimize the effects of atmospheric extinction and dispersion.
We imposed a $\sim 20 \arcsec$ non-repeating dithering pattern between 
consecutive images to minimize the effects of bad pixels and to allow 
for the creation of a sky flat by combining unregistered images
(super-sky flat).  Dithering by this amount had the effect of reducing 
the field size common to all images by $\sim 1 \arcmin$ in both 
dimensions.  The integration times were varied depending on filter, 
observing conditions, and telescope/instrument configuration in an
attempt to achieve our targeted 1$\sigma$ magnitude sky fluctuation
depths of $u'_{AB}$ = 29.0, B$_{AB}$, V$_{AB}$, and R$_{AB}$ = 28.5,
and I$_{AB}$ = 27.8 with a seeing element goal of $< 1\arcsec$.  The 
average observed 1$\sigma$ magnitude sky fluctuation depths per seeing 
element were $u'_{AB}$ = 29.1, B$_{AB}$ = 28.7, V$_{AB}$ = 28.6,
R$_{AB}$ = 28.7, and I$_{AB}$ = 27.6 with an average seeing element of 
$0.\arcsec97\pm0.2$.  Since the seeing for most fields was 
consistent, any $u'$, B, V, or I image with a seeing FWHM differing 
from the corresponding R filter image by more than $0.\arcsec2$ 
was not used.  This resulted in compatible seeing for
each field in the final stacked images.  
%Due to time constraints, instrument failures, and variable seeing 
%conditions, 
For the three fields LBQS0056+0125, PKS0336--017, and PC1643+4631A, the 
stacked images consisted of only four filters.  This was due to these 
fields having either a broad seeing FWHM from poor seeing or non-science 
grade images in a particular filter.  These fields were only used in specific 
photometric selection and LBG candidate density results.  

The data were reduced using standard $IRAF$ and $IDL$ routines. 
Images were flat-fielded using super-sky flats whenever possible. 
In a few cases, the instrument condition or the particular night's 
agenda forced the use of a twilight flat or dome flat.  Final images 
were combined with an in-house package that inversely weights each 
image by the measured variance, scales by the exposure time, and 
creates $1\sigma$ error images based on Poisson statistics.
Photometric calibrations were calculated for each night by observing 
multiple photometric standard stars \citep {l92} in each 
filter over a full range of airmass. Conventional $IRAF$ tasks, such 
as $PHOT$ and $FITPARAMS$ were employed to solve zero-point magnitudes, 
airmass correction coefficients, and appropriate color correction 
coefficients for the photometric data.   Typical rms magnitude errors 
were $\sim 0.04, \sim 0.02,$ and $\sim 0.01$ respectively.  In the 
few cases where a unique solution was not found, we adopted fixed values 
for the airmass correction coefficients.  These values are: 0.56,
0.35, 0.17, 0.13, 0.07 for $u'$BVRI filters respectively at Palomar 
and 0.41, 0.19, 0.12, 0.11, and 0.07 for $u'$BVRI filters respectively 
at Keck. Finally, the photometric zero-points were corrected for Galactic 
extinction using $E(B-V)$ values measured from the far-IR dust emission 
maps of \citet {schlegel98}.  The $E(B-V)$ values for each field
are given in Table~\ref{imaging}.  To correct our fields for Galactic 
extinction, we adopted the following relations interpreted from \citet 
{ccm89}: $$A(u') = 4.8 ~E(B-V), ~A(B) = 4.1 ~E(B-V), ~A(V) = 3.1 ~E(B-V)$$
$$A(R) = 2.3 ~E(B-V), ~A(I) = 1.5 ~E(B-V)$$

%------------------------------------------------------------------------

\section{PHOTOMETRY}

We defined our source detection catalogs in the R-band 
(R$_s$ in 2 fields) after considering the spectral profile of $z\sim3$
LBGs and the imaging sensitivity of the $LRIS$ and $COSMIC$ cameras.  
An additional consideration was that $\sim30\%$ of the PSS1432+3940 
and JVAS2344+3433 fields (imaged with the larger FOV of the $COSMIC$
camera) had only $u'$BR filter coverage.  The sources in all nine 
QSO fields were identified with the software package $SExtractor$ 
(v.2.1.6) created by \citet {ba96}.  We varied the `{\tt $\sigma$ 
detection threshold}' and `{\tt minimum pixel area}' input parameters 
by comparing the output segmentation map of each stacked R image with 
that of the segmentation map of the negative stacked R image to 
minimize spurious source detections. Typical $SExtractor$ values were 
1.6 $\sigma$ detection threshold and 6 pixel minimum area. An 
in-house code was used to calculate the 
%In nearly all fields, the seeing between the five
%filters was similar enough to allow magnitude measurements using the 
%pixel definitions identified by the $SExtractor$ segmentation map of 
%the stacked R-band image.   
magnitudes and errors of each object detection in each filter.  This 
code determines the sky flux by taking the bi-weight \citep {b90} 
of the sky in an annulus around each object containing at least 
10$^3$ pixels not flagged as an object in the segmentation map created 
by $SExtractor$.  Magnitude errors were derived from variance in the sky 
flux and the systematic uncertainty in the photometric calibration. 

\begin{figure}
\begin{center}
\scalebox{0.36}[0.28]{\rotatebox{90}{\includegraphics{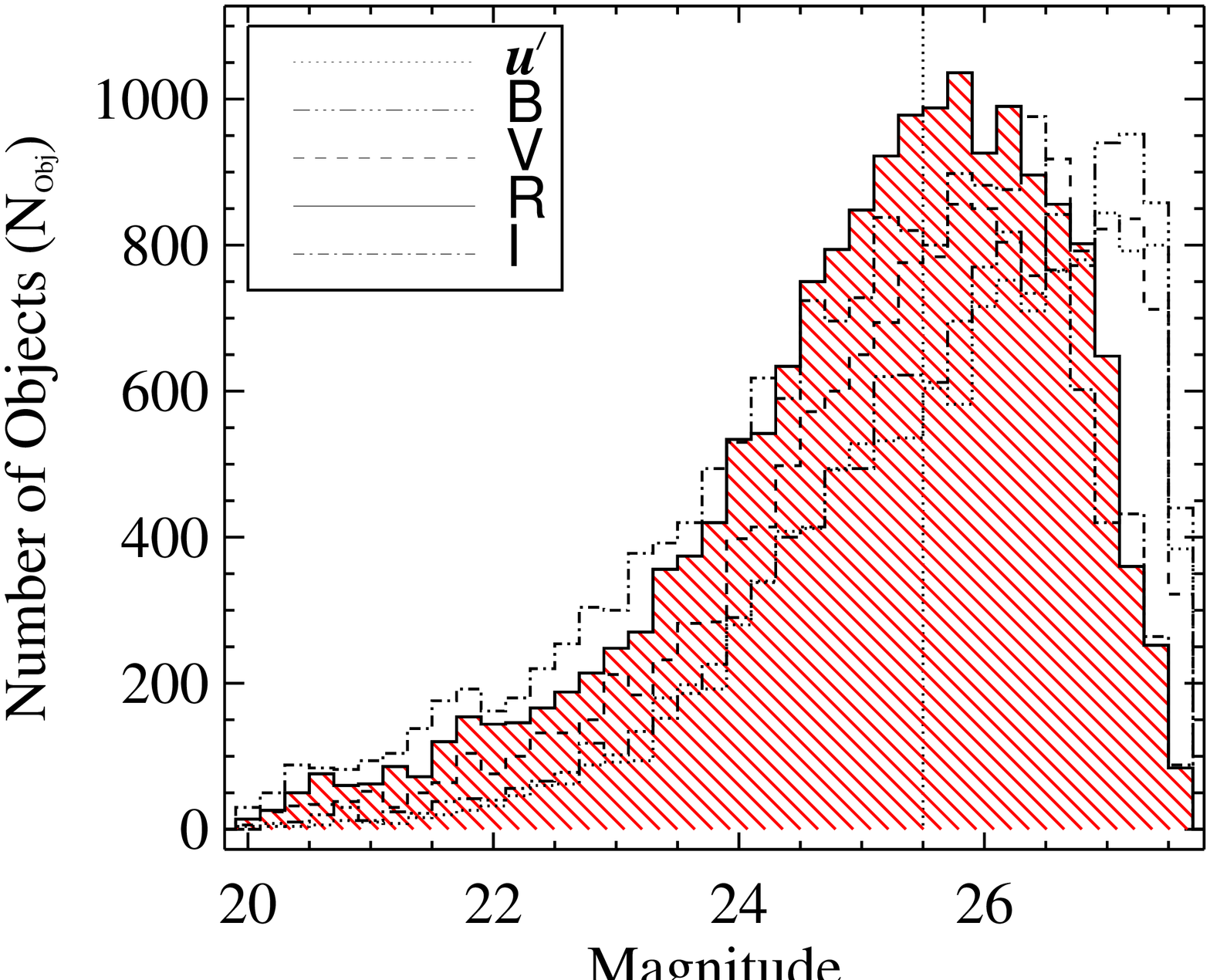}}}
\scalebox{0.36}[0.28]{\rotatebox{90}{\includegraphics{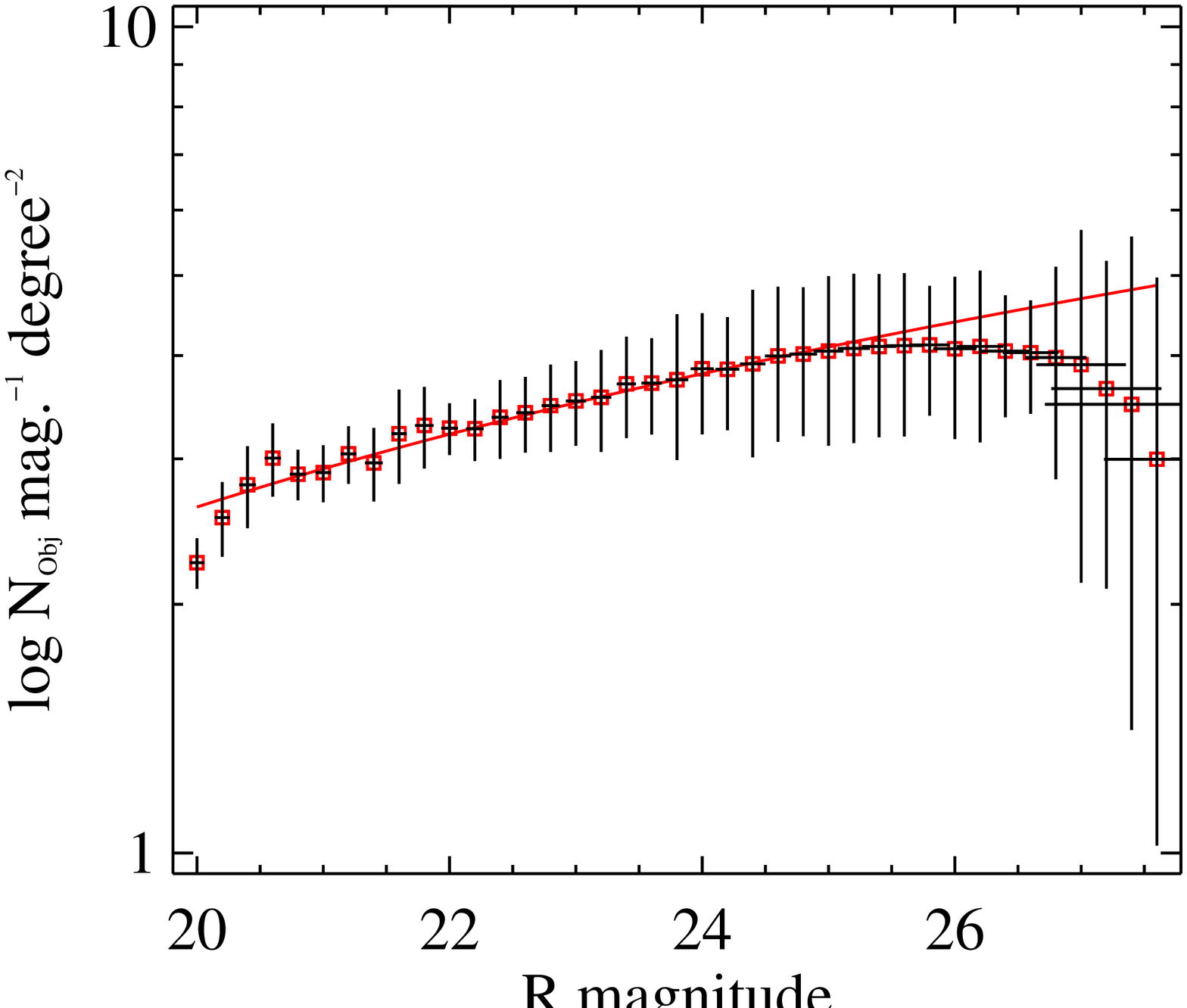}}}
\scalebox{0.36}[0.28]{\rotatebox{90}{\includegraphics{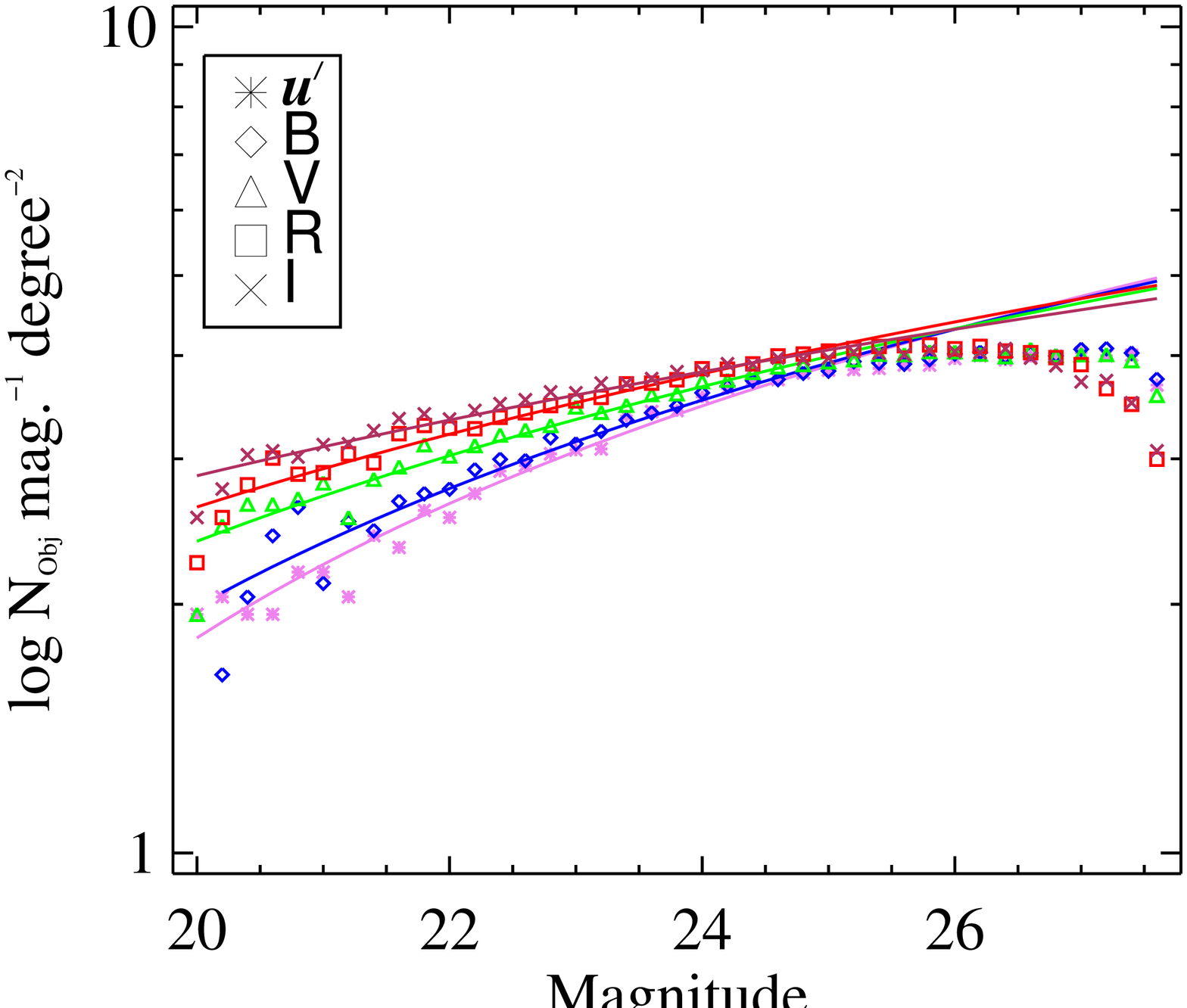}}}
\caption{\small Number counts versus magnitude in five fields with
$u'$BVRI imaging.  This method is used as a check for
photometric completeness.  The upper histogram shows the number of 
source detections versus R magnitude (hatched region) binned in
$\Delta$R$=0.2$.  Other filters are plotted versus their respective
magnitude with line styles indicated in the legend.  The spectroscopic 
magnitude limit of $LRIS$ is denoted by a vertical dotted line at
R$=25.5$.  The central plot displays a logarithmic fit (solid line) to
the number counts versus R magnitude.  The lower plot displays the 
logarithmic fit to all five filters for direct comparison.  In this
plot, the error bars have been removed for clarity.  The slope of the 
logarithmic fit flattens for consecutively redder filters. 
\label{maghist}} 
\end{center}
\end{figure}

A goal in imaging the survey fields was to achieve photometric 
completeness to the $LRIS$ spectroscopic limiting magnitude of
R$\sim25.5$.  The number of objects detected in our survey versus
their R magnitude is shown in Figure~\ref{maghist}.  The upper
histogram displays the number counts of sources for each filter in 0.2
magnitude bins for the five fields with $u'$BVRI imaging.  A
logarithmic fit to the binned number counts to R$=25.5$ is used to 
estimate the completeness of our sample and is shown in the central 
plot.  The lower plot displays the photometric completeness for all 5
filters. These fields are complete to R$\sim$25 and are $\ge80\%$ 
complete to R=25.5 using this form of analysis.  From the deep galaxy 
survey of \citet {s03}, the average number of faint galaxies in the
apparent magnitude range of $22.5<{\cal R}<25.0$ was found to be
25.9 with a field-to-field scatter of $\pm1.9$ arcmin$^{-2}$.  We
report an average faint galaxy number of $24.9\pm3.5$ arcminute$^{-2}$ 
that meet our source identification criteria over the same magnitude 
range in the R-band.

To facilitate the design and milling of reliable spectroscopic
multislit masks, accurate astrometry was performed over the full area
of each field using USNO A2.0 and USNO B1.0 \citep {monet98,monet03}
astrometric catalogs. We then used the PAL\_COORDS 
routine\footnote{Available by anonymous ftp at: 
ftp.astro.caltech.edu/palomar/pal\_coordinates/pal\_coordinates.doc} 
to determine the astrometric solution for both $LRIS$ and $COSMIC$ 
cameras.  Typical position errors of $\sim 0.\arcsec 2$ were reported
by this method and confirmed when fine aligning the slitmasks on sky.

%------------------------------------------------------------------------

\subsection{Photometric Color Selection}

We chose to identify the color of all LBG candidates in a consistent 
manner from their coarse spectral features.  We searched for star 
forming galaxies having a strong continuum long-ward of rest-frame 
Lyman $\alpha$ (1215\AA) and a flux decrement short-ward of the 
rest-frame Lyman limit (912\AA).  This decrement is due to absorption 
by H I optically thick at the Lyman limit which is intrinsic to the 
galaxy and present in foreground systems.  At $z\sim3$, these spectral 
characteristics are redshifted to optical wavelengths.  The
photometric selection of $z\sim3$ LBG candidates in color-color space 
has been spectroscopically verified with ground-based telescopes 
\citep{s96} using $U_n G {\cal R}$ broad band filters and with the 
Hubble Space Telescope [\citet{st96} and \citet{l97}] using similar
bandpasses.  Here we use five widely available filters, $u'$, B, V, R, 
and I, to formulate our LBG candidate selection criteria. 

\begin{figure}
\begin{center}
\scalebox{0.55}[0.5]{\rotatebox{90}{\includegraphics{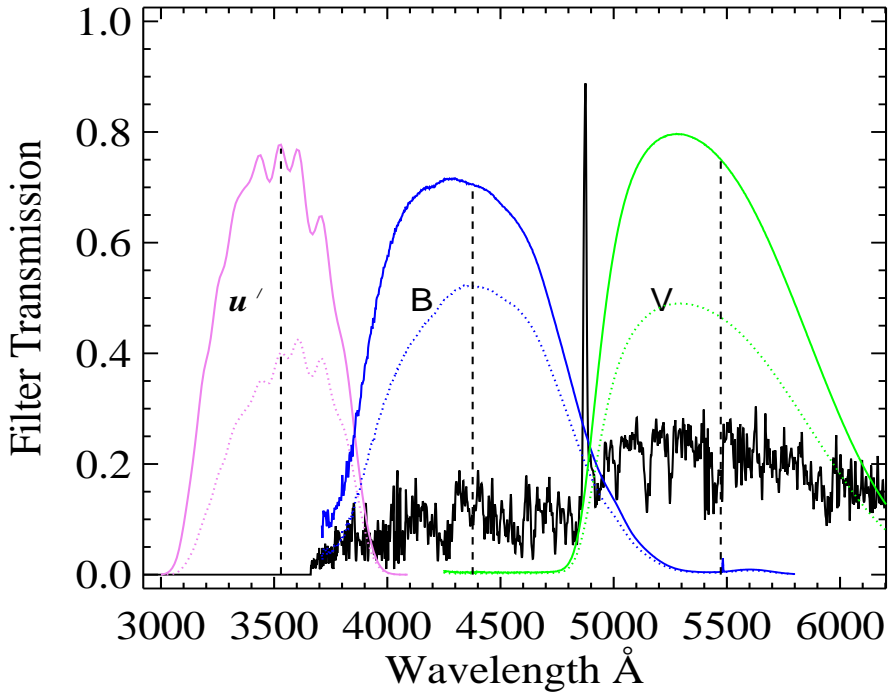}}}
\caption{\small Graphical depiction of the Lyman break photometric
selection.  The filter transmissions of the three filters ($u'$BV) are 
shown with their transmissions before (solid curves) and after the 
correction for CCD quantum efficiency and atmospheric attenuation
(dotted curves).  Vertical dashed lines indicate the effective
wavelengths of each filter.  Overlayed is a spectrum of a bright LBG
from the data with its wavelength array shifted to represent an 
arbitrary LBG at $z=3.0$. Short-ward of 912\AA~in this spectrum is 
shown as zero flux.  The placement of the break in the continuum of 
the galaxy at the rest-frame Lyman limit (912\AA) and at rest-frame 
Lyman $\alpha$ can be seen with respect to the filter responses.  
This particular LBG displays a Lyman $\alpha$ emission spike (near 
4900\AA).  Our color criteria was chosen to detect a distribution of 
objects that have the Lyman $\alpha$ break centered at $z=3.0$ causing 
their Lyman limit break to be redshifted to within the passband of the 
$u'$ filter. \label{ubvLBG}} 
\end{center}
\end{figure}

\begin{figure}
\begin{center}
\scalebox{0.35}[0.42]{\rotatebox{90}{\includegraphics{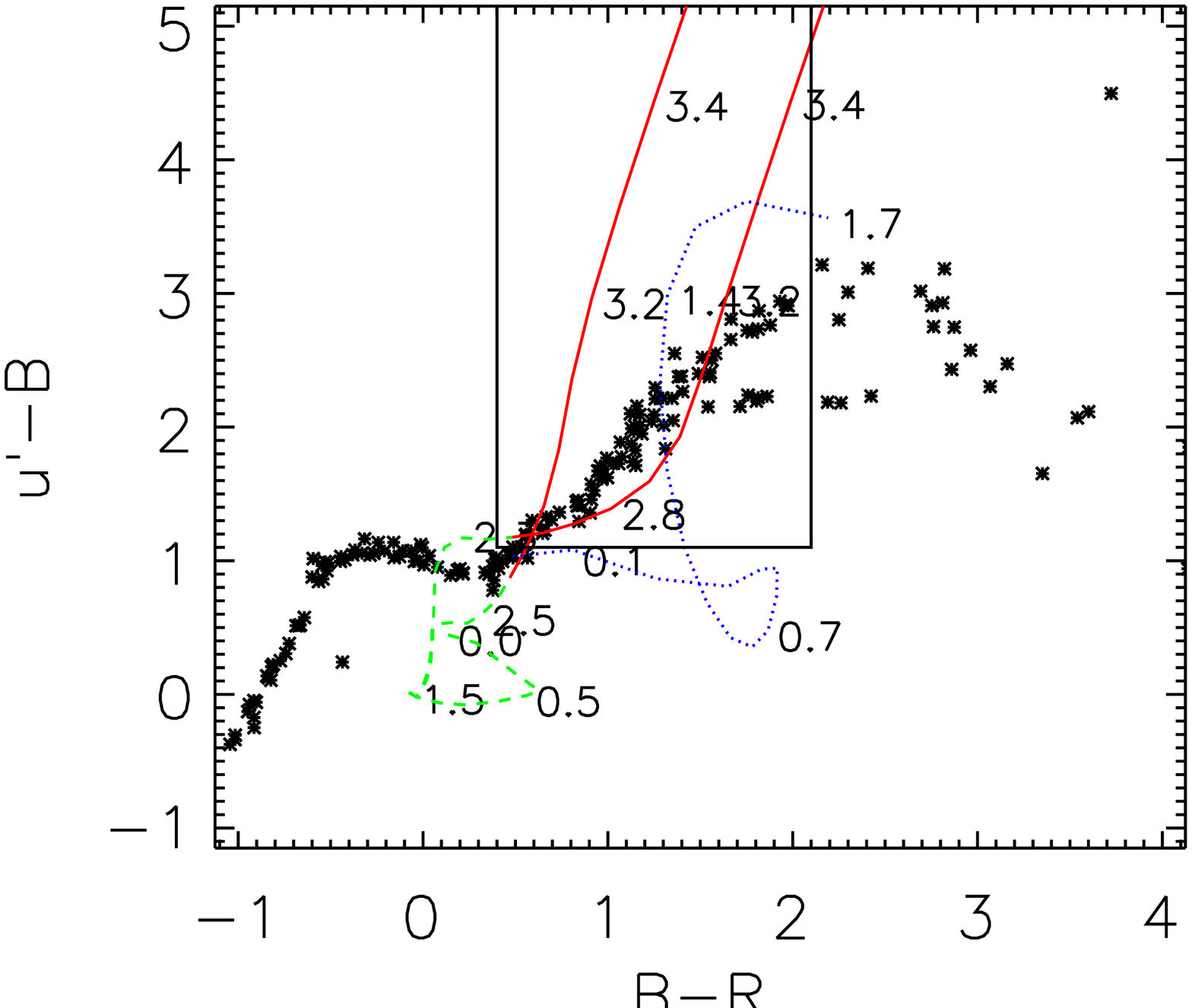}}}
\scalebox{0.35}[0.42]{\rotatebox{90}{\includegraphics{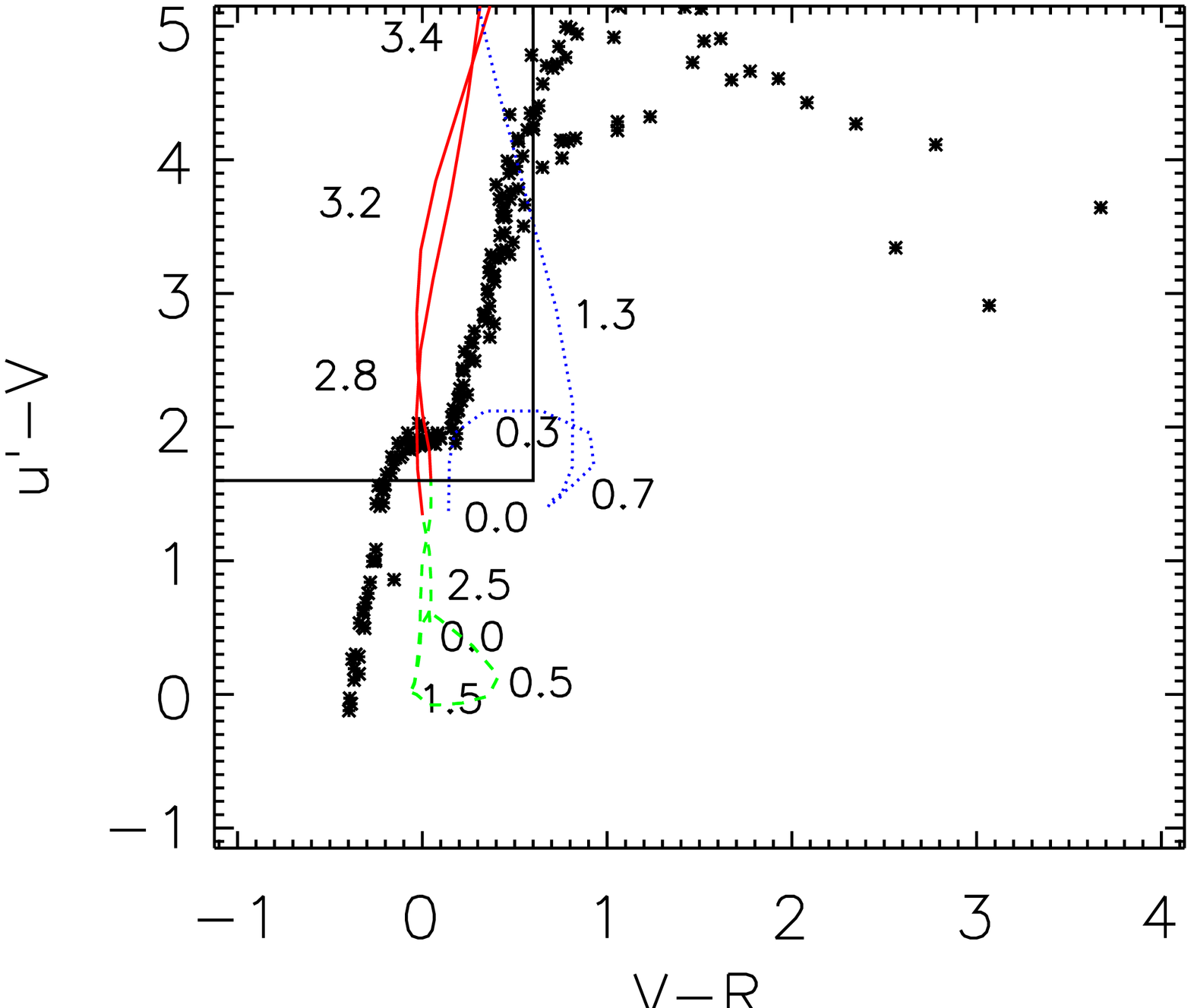}}}
\scalebox{0.35}[0.42]{\rotatebox{90}{\includegraphics{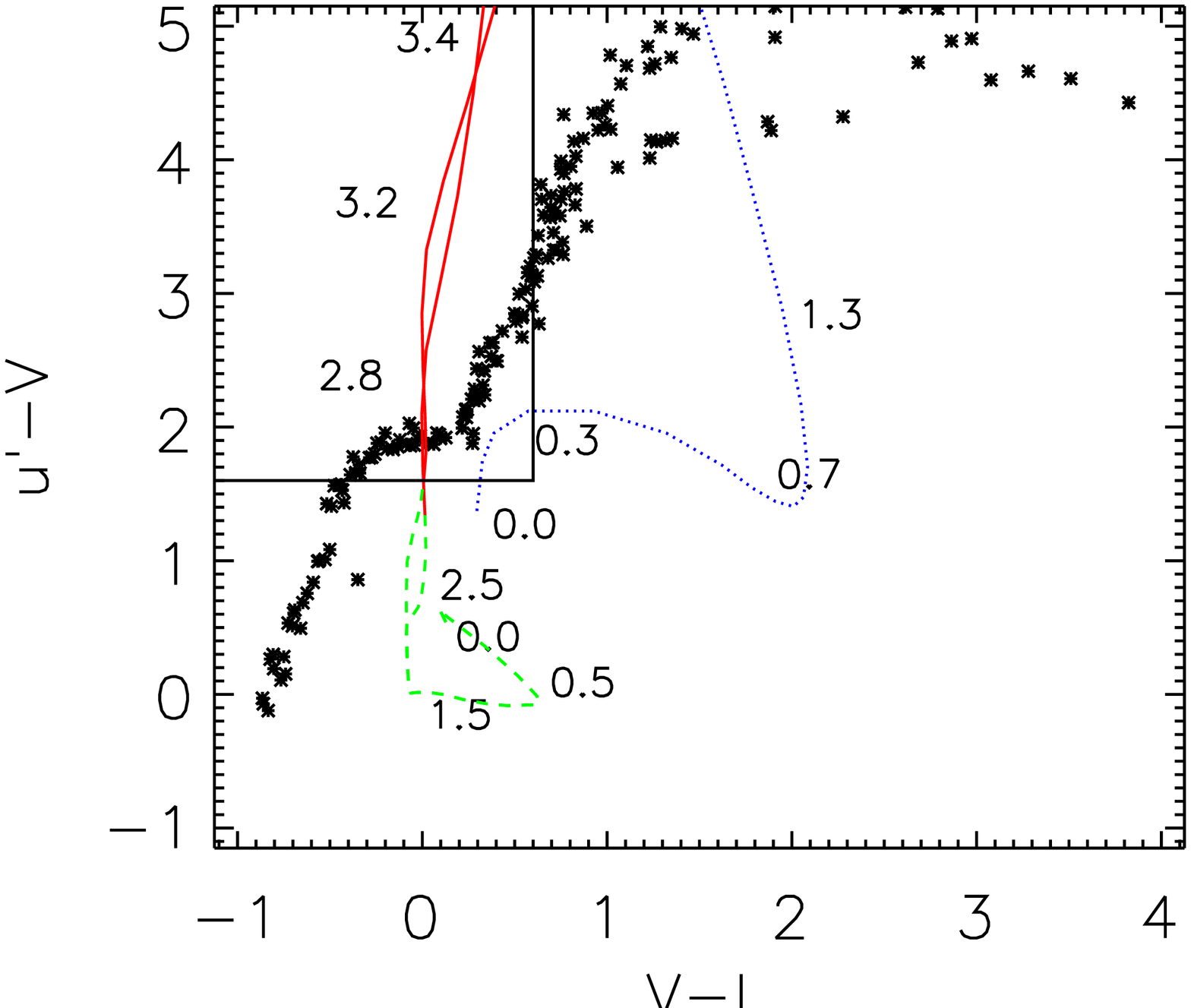}}}
\caption{\small Color-color plots showing the expected colors of a
starbursting galaxy (dashed green curve) and an elliptical galaxy
(dotted blue curve) as they are evolved from $z=0$.  The starbursting 
galaxy is indicated by two solid red curves from $2.5<z<3.5$.  The 
separate red curves indicate decrements in the continuum of the 
galaxy caused by a reasonable range of Lyman $\alpha$ forest absorption. 
Star symbols illustrate the colors of Galactic stars from the Gunn and 
Stryker (1983) catalog.  The LBG candidate selection regions are the 
boxed areas in the upper-central or upper-left portions of the plots.  
Low redshift reddened elliptical galaxies and low-mass Galactic stars 
are the main source of interlopers. \label{theoplots}}
\end{center}
\end{figure}

Galaxies meeting our selection criteria exhibit virtually no flux in
the $u'$-bandpass, a varying flux decrement in the B-bandpass due to 
the Lyman $\alpha$ forest, and relatively constant flux in the VRI 
bandpasses.  Figure~\ref{ubvLBG} displays a graphical representation
of this in the $u'$, B, and V filter bandpasses.  The placements of
the Lyman $\alpha$ and Lyman limit breaks in the continuum are readily 
apparent.  We derived theoretical galaxy colors using six star-forming 
galaxy templates \citep{y00} and simulated the color-color evolution 
of these galaxies to high redshift.  Figure~\ref{theoplots} plots the 
expected $(u'-B)$ vs.~$(B-R)$, $(u'-V)$ vs.~$(V-R)$, and $(u'-V)$ 
vs.~$(V-I)$ colors of a starbursting galaxy (dashed green curve) and
an elliptical galaxy (dotted blue curve) as they are evolved from 
$z=0$. Star symbols are used to denote the colors of Galactic stars
from the \citet{gs83} catalog. From approximately $2.5<z<3.5$ in each 
plot, the starbursting galaxy is indicated by two solid red lines.  
The separate red lines indicate the decrements in the continuum of 
the galaxy caused by a reasonable range of Lyman $\alpha$ forest 
absorption \citep{madau95}.  The expected colors of these objects were 
used to determine the appropriate color criteria that would
efficiently detect Lyman break candidates at $z\sim3$ and reject low 
redshift objects.  The color selection regions are shown as the boxed 
regions in the upper-central and upper-left portions of the plots. 

Objects were selected as LBG candidates that satisfied the following 
constraints:
\begin{equation}
(u'-B)_{AB} > 1.1,
\end{equation}
\begin{equation}
(u'-V)_{AB} > 1.6,
\end{equation}
\begin{equation}
0.6 < (B-R)_{AB} < 2.1,
\end{equation}
\begin{equation}
(V-R)_{AB} < 0.6,
\end{equation}
\begin{equation}
(V-I)_{AB} < 0.6,
\end{equation}
\begin{equation}
20.0 < R_{AB} < 25.5.
\end{equation}

%\subsection{Spectroscopic Slitmasks}

LBG candidates for the first two spectroscopic slitmasks of the 
field JVAS2344+3433 were determined using the $(u'-B)$ vs. $(B-R)$ 
color-color plot only (constraints $4,6, \&~9$).  This was necessary 
prior to complementary V and I imaging at Keck and remains
the color selection criteria for the regions of fields PSS1432+3940 
and JVAS2344+3433 having only $u'$BR imaging.  LBG candidates in fields 
where the images in all five filters were usable were required to meet
the full color criteria (constraints $4-9$).  Constraints 4, 5, and 6 
were imposed to reflect the relative discontinuities in the continuum 
at rest-frame 912\AA~and 1215\AA~from the continuum at rest-frame 
$\sim$1300\AA.  Constraints 7 \& 8 were used to eliminate a large
number of low redshift reddened elliptical galaxies and Galactic
stars.  The R magnitude (equation 9) of all objects was  constrained
on the bright end (R$_{AB} > 20.0)$ to avoid saturated
sources in the combined images and constrained on the faint end
(R$_{AB} < 25.5)$ to only include objects bright enough to meet the 
spectroscopic limit of $LRIS$ on Keck.  The imposition of color 
constraints similar to the ones mentioned above selects the majority 
of the galaxies at $z\sim3$ from field objects but does not allow for 
the selection of $z\sim3$ objects with excessive intrinsic far-UV 
extinction.  Therefore, this survey is not complete to all possible 
types of galaxies at $z\sim3$, but does include a significant number 
of galaxies to provide meaningful statistics toward the goal of 
cross-correlating LBGs with DLAs.  

\begin{figure}
\begin{center}
\scalebox{0.43}[0.35]{\includegraphics{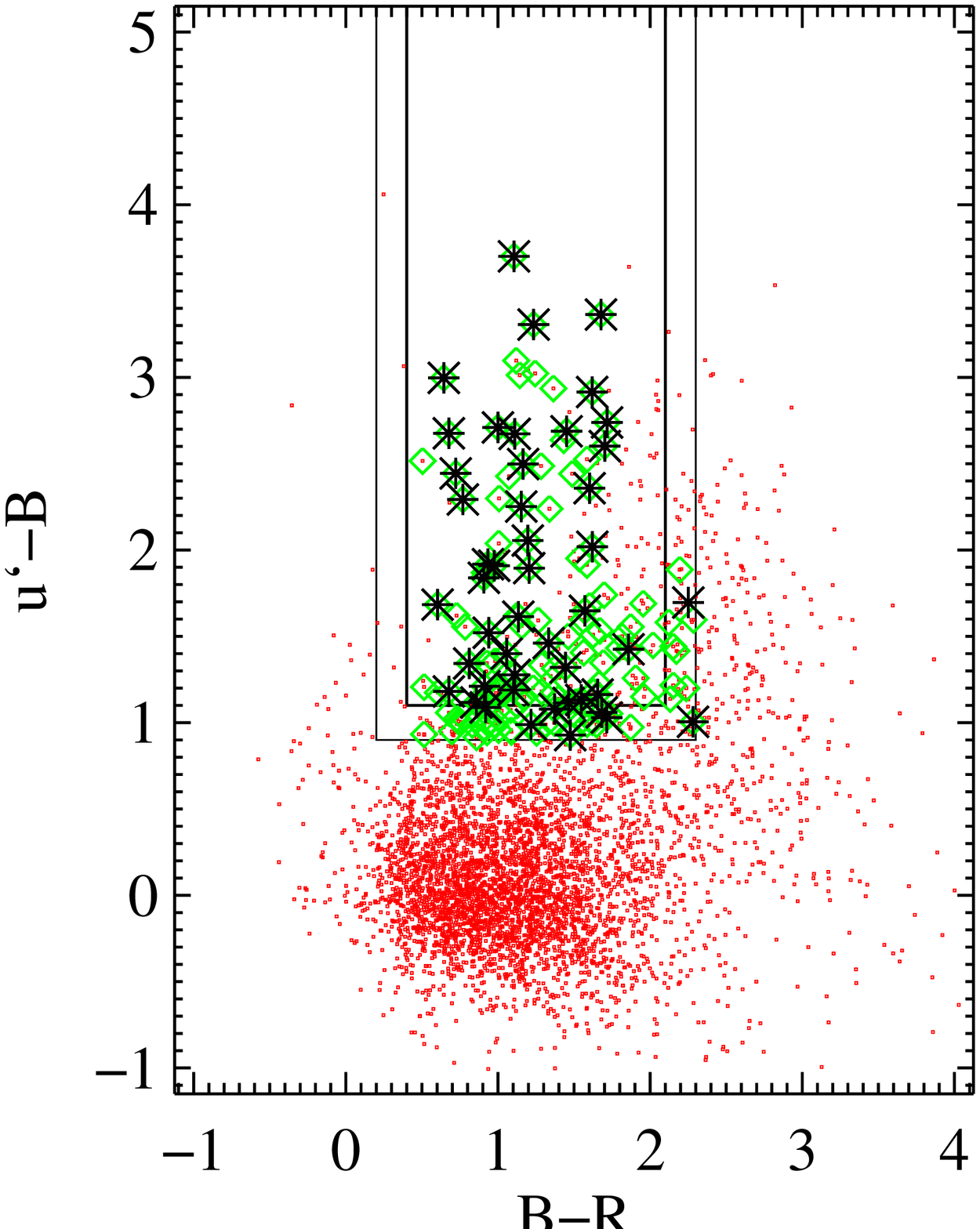}}
\scalebox{0.43}[0.35]{\includegraphics{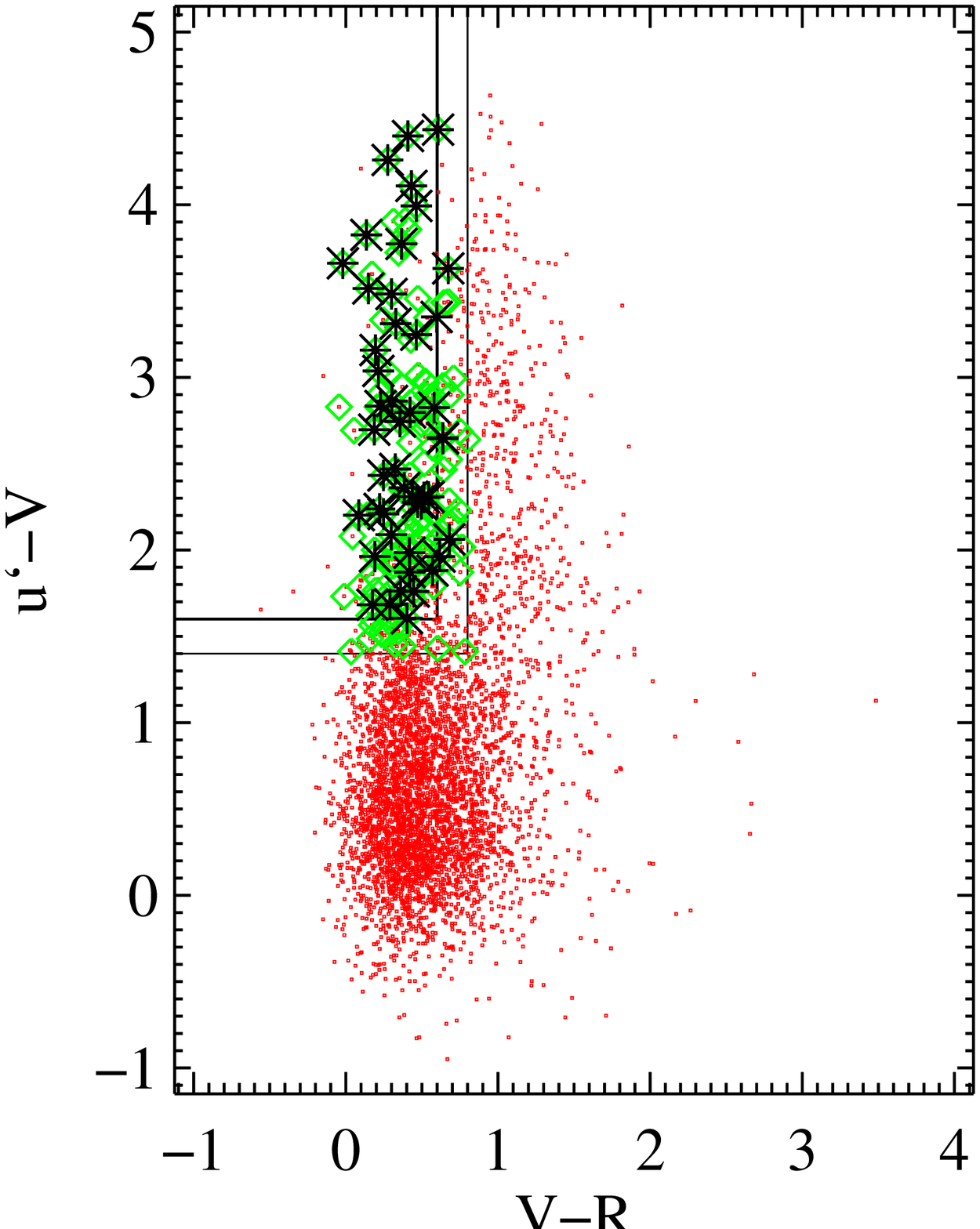}}
\scalebox{0.43}[0.35]{\includegraphics{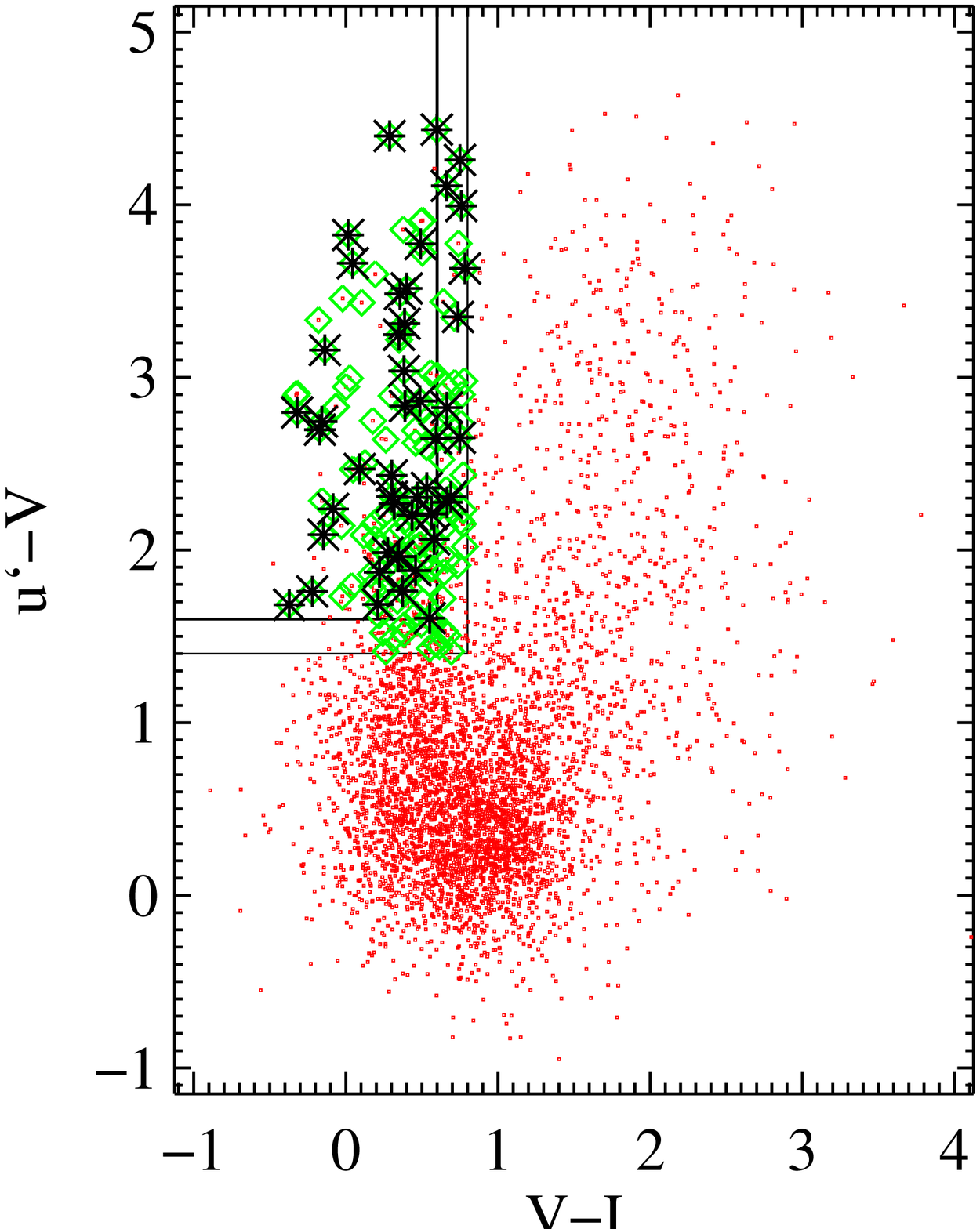}}
\caption{\small Color-color plots of the data from the fields with 
$u'$BVRI color selection. All $20.0<$ R $<25.5$ objects are
denoted by red points. Those objects meeting all $u'$BVRI color 
criteria are indicated by green diamonds and reside in the boxed 
selection regions in the upper central, or upper left portions of 
the plots.  Black asterisks denote objects meeting all $u'$BVRI 
color criteria and have secure $z > 2$ redshifts.  The boundary 
regions surrounding the selection regions reflect a relaxation of 
the color selection criteria by 0.2 magnitudes to compensate for 
photometric errors.  Most of the objects selected are not detected 
in the $u'$-band, therefore the values shown are the 2$\sigma$ lower 
limit to their $(u'-B)$ and $(u'-V)$ colors. \label{dataplots}}
\end{center}
\end{figure}

Objects meeting the full $u'$BVRI color criteria (constraints $4-9$)
were given the highest priority as slitmask targets.  Objects meeting 
only $u'$VRI and $u'$BRI color criteria were given successively lower 
priorities.  To compensate for photometric errors due to both
statistical and systematic uncertainties, objects within 0.2
magnitudes outside of these constraints were also identified as LBG 
candidates and given respectively lower priorities.  LBG candidates
within a separation of $\sim10\arcsec$ in radius from a QSO sight-line
were marked and chosen over other objects that were in conflict in the
dispersion direction of the particular slitmask in an effort to identify
the galaxy responsible for the damped absorption.  A separation of
$\sim10\arcsec$ represents the maximum impact parameter of an absorber 
inferred from absorption line statistics \citep {slw00}.  In some
cases, the conflicting LBG candidates were placed on subsequent
slitmasks.  No other factor contributed to the choice of candidates
for spectroscopy except the physical constraints of the slitmask
itself, as described in $\S$~4.  The three 
color-color plots in Figure~\ref{dataplots} display the data from 5 
fields (PSS0808+5215, PSS0957+3308, BRI1013+0035, PSS1057+4555, and
PSS1439+3940) having full $u'$BVRI color selection. In this figure, 
all $20.0<R<25.5$ objects in the five fields are shown as red points.
Green diamonds are overlayed to indicate those objects that met the 
imposed $u'$BVRI color selection criteria (including objects within 
0.2 magnitudes) and the black asterisks denote the objects that have 
secure $z > 2$ spectroscopic redshift identifications.

Overall, a total of 796 objects were photometrically selected using
these color criteria.  In the four fields where all five filters 
could not be utilized, candidates were chosen that met the color 
criteria of the usable filters.  Firstly, the initial slitmasks for 
the JVAS2344+3433 field included LBG candidates chosen by $u'$BR color 
selection prior to V and I filter imaging.  Secondly, the LBG candidates 
in both the LBQS0056+0125 and PC1643+4631A fields were chosen using 
$u'$VRI color selection.  An instrument failure resulted in non-science 
grade B filter images for the LBQS0056+0125 field and time constraints 
forced the full bandpass of the B filter to be compromised by a dichroic 
centered at 4600\AA~for the PC1643+4631A field.  Lastly, the LBG 
candidates in the PKS0336--017 field were chosen using $u'$BRI color 
selection due to complications from merging a broad and shallow FWHM 
of 1.\arcsec7 in the V filter.  All LBG candidates in this field were 
required to meet an additional color criterion of: 
\begin{equation}
0.6 < (B-I)_{AB} < 2.1.
\end{equation}

Table~\ref{colors} displays the area common to all filters for each 
targeted field and summarizes the number of LBG candidates detected 
using $u'$BR(I), $u'$VRI, and $u'$BVRI color selection criteria and
the resulting $u'$VRI and $u'$BVRI LBG number densities.  In the five 
fields with $u'$BVRI photometry, the two-dimensional LBG color
candidate density is 0.79 arcmin$^{-2}\pm0.25$ arcmin$^{-2}$.  
Comparison of the average LBG candidate density with that of similar 
surveys can be used as a photometric selection check.  From \citet
{s03}, the LBG candidate density using $U_n G {\cal R}$ I filters is 
1.6 $\pm0.5$ arcmin$^{-2}$ from 2347 selected objects in 17 fields.  
Our average LBG candidate density using the relatively comparable 
$u'$VRI filters is consistent at 1.45 $\pm0.66$ arcmin$^{-2}$ from 365 
objects in 6 applicable smaller fields.  Although the $u'$VRI criteria
select a larger number of objects than the $u'$BVRI criteria, we find 
the $u'$VRI method to be less efficient in resulting $z>2$ 
spectroscopic confirmations (see $\S$~5).

% **  mention tentative DLA in emission results
%Using the photometric color criteria above (constraints $4-9$), we 
%generated lists of LBG candidates and assigned a numerical priority 
%code to each depending on their colors.  An important additional goal
%of our survey is to identify galaxies in emission responsible for the 
%damped Ly$\alpha$ absorption in the spectra of the targeted QSOs.  A 
%separation of $\sim10\arcsec$ in radius from a QSO sight-line
%represents the maximum impact parameter of an absorber inferred from 
%absorption line statistics \citep {slw00}. Objects that met the 
%photometric color and separation criteria were given the highest 
%priority code.  

%------------------------------------------------------------------------

\section{SPECTROSCOPIC DATA}

To optimize spectroscopic slitmask design, objects on the LBG
candidate lists were visually inspected in the two-dimensional images
to eliminate false detections (e.g. satellite trails, diffraction
spikes, asteroid tracks, and CCD defects).  LBGs have a half-light 
radius of $<0.\arcsec5$ from Hubble Space Telescope images \citep
{st96,l97,g00} and should appear nearly point-like with the resolution 
of $LRIS$ and $COSMIC$.  Therefore, several objects were rejected if
they were overly extended [although some $z\sim2$ objects appear 
to be extended \citep{e04}]. Each slitmask was then designed to 
include the highest priority objects and to have empty regions filled 
with objects of successively lower priorities.  We chose to use
6$\arcsec$ minimum slitlet lengths to maximize the number of objects 
observed without compromising sky subtraction.  Slitmasks were 
designed using the software AUTOSLIT 3.07 \footnote{Autoslit and
support files are available at
http://astro.caltech.edu/$\sim$pls/autoslit/ and 
ftp://astro.caltech.edu/users/pls/autoslit/}. Typically, each slitmask 
was able to include $25-35$ objects which resulted in an average 
slitlet length of $\sim16 \arcsec$ for each candidate.

Multislit spectroscopic observations of the LBG candidates were
obtained between 2000 November through 2004 February, using both
the red arm and new blue arm of $LRIS$ on the Keck I telescope.  We
used the 300 line mm$^{-1}$ grating (red side) and the 300 line
mm$^{-1}$ grism (blue side), both blazed at 5000\AA~in first order.  
The red side grating has a dispersion of 2.47\AA~pixel$^{-1}$ over
the $2048\times 2048$ CCD and was tilted to allow a minimum
wavelength range of $3800-7000$\AA~for each slitlet regardless of
their placement on the mask.  The blue side grism has a dispersion of 
1.43\AA~pixel$^{-1}$ ($15\mu m$ pixel size) over two Marconi high
quantum efficiency (QE) blue sensitive $2048\times 4096$ CCDs.  The 
grism is fixed and designed to allow a possible instrumental
wavelength range of $2210-8060$\AA~for multislit spectroscopy.  This 
results in an observational minimum wavelength coverage from the 
atmosphere cut-off at $\sim3200$ to $\sim$7500\AA~for each slitlet.  
We designed the widths of the slitlets on most slitmasks to be 
$1.\arcsec 5$ to account for errors in the astrometry, atmospheric 
dispersion over long integrations, and moderate seeing.  Since these
are sky-dominated objects, when possible, we designed alternate
slitmasks with narrower slitlet widths $(1.\arcsec0-1.\arcsec3)$ to 
accommodate good seeing and observations at low airmass through the 
meridian.  The spectral resolution evaluated from night sky emission 
lines was $\sim$12\AA~FWHM and, because Lyman break objects are near
point sources, typical object resolution was seeing limited and
resulted in an effective spectral resolution of $\sim$9\AA~FWHM.  

The total exposure time for each slitmask was 7200 seconds when using 
the red arm and $\sim3600-4800$ seconds when using the blue arm of
$LRIS$.  The higher throughput, lower read-out noise, and higher CCD 
quantum efficiency of the blue arm permitted considerably shorter 
exposure times to achieve equivalent magnitude depths as compared to 
the red arm.  Each slitmask was observed in a series of consecutive 
$1200-1800$ second exposures with small dithers
$(0.\arcsec5-1.\arcsec0$) along the direction of the slits to sample 
different parts of the CCD, minimize fringing effects, avoid hot
pixels, and provide better sky subtraction options. The details of the 
spectroscopic observations are summarized in Table~\ref{spectro}.  The 
slitmasks were designed to be observed at position angles equaling the 
parallactic angle determined mid-way through their scheduled time of 
observation.  In some instances (to help compensate for uncertain
weather and instrument conditions), two masks were designed with
position angles reflecting two different times of possible
observation.  This practice minimized wavelength dependent flux loss 
through the slit due to atmospheric dispersion as the slitmasks were 
only observed within 30 minutes of their scheduled time.  Calibration 
exposures were taken either before, mid-way, or after the science 
exposures to minimize the effects of instrument flexure.  Flat field 
exposures were obtained using an internal halogen lamp.  Internal arc 
lamp exposures used for wavelength solutions were taken with a
combination of Cd, Zn, Hg, Ne, Ar, Kr, and Xe, depending whether the 
science exposures were taken with the red or blue arm of LRIS.  For 
proper flux calibration, two or more spectroscopic standard stars 
\citep {m88} were observed each night at different airmass to
replicate the range of airmass sampled by the slitmasks during the 
course of the night.

We used IRAF and in-house IDL scripts to perform the initial reduction 
of the CCD data including overscan correction, gain normalization, 
trimming, bias subtraction of all images.  We used the
$IRAF$ routine {\tt bogus.cl} \footnote{bogus.cl is available at:
http://zwolfkinder.jpl.nasa.gov/$\sim$stern/homepage/bogus.html} 
to handle the division of each full two-dimensional multislit image 
into separate two-dimensional slitlet images of each object and for 
the application of calibration frames.  The {\tt bogus.cl} routine calls 
$IRAF$ tasks for flat fielding the individual slitlet images by their
corresponding continuum-normalized portion of the flat field image.
Background sky subtractions were determined by using low-order polynomial
fits appropriate to the length of each slitlet to background sky
sections.  The routine also performs fringe subtraction, identifies 
cosmic rays morphologically, and shifts and stacks the spectra. 
Additionally, {\tt bogus.cl} creates two-dimensional slitlet images
during selected steps for later inspection and analysis.  We chose to 
operate interactively on all spectra to achieve our goals considering 
the faint magnitude of typical objects in our survey.

We traced the reduced two-dimensional stacked spectra using $IRAF$ 
spectroscopic analysis routines and fitted low-order (usually
first-order) polynomials to subtract any remaining background sky 
offset. Corresponding arc lamp traces of each object were extracted 
in the identical manner as the object.  Wavelength solutions were 
obtained on the extracted arc lamp spectra using low order polynomials 
with residuals $<$ 1\AA.  Further refinement to the wavelength
solution was performed by comparing the arc lamp wavelength solution
to multiple night sky emission lines corresponding to each object and 
applying the calculated sub-pixel offset.  Each spectrum was then 
corrected for extinction, flux calibrated using sensitivity functions 
obtained from the corresponding spectrophotometric standard stars, and 
corrected to vacuum wavelengths.  

%------------------------------------------------------------------------

\subsection{Spectroscopic Identification and Analysis}

LBGs display identifiable rest-frame UV spectral features from 
912\AA~to $\sim$1700\AA.  For galaxies at $z\sim3$, these features are
shifted to optical wavelengths from $\sim3200-7500$\AA. The objects in 
this survey were identified mainly using Lyman $\alpha$ and
interstellar emission and absorption features described by \citet
{p00}, \citet {a03}, and \citet {aes03}.  These features have typical 
equivalent widths of $\sim5-10$\AA~and can be relatively easy to
identify even though the spectra have a signal-to-noise ratio of only
a few.  

We chose to separate the LBGs detected by this survey into two simple 
populations: (a) emission-identified LBGs -- galaxies showing Lyman 
$\alpha$ in emission and (b) absorption-identified LBGs -- galaxies 
showing Lyman $\alpha$ in absorption.  Confirmation for all LBGs
involved the individual search through each continuum for a possible 
decrement due to the Lyman $\alpha$ forest short-ward of the Lyman 
$\alpha$ feature at rest-frame 1215\AA, a second decrement (where 
spectral coverage allowed) due to Lyman limit absorption short-ward of 
rest-frame 912\AA~, and the identification of expected interstellar 
absorption and/or emission features.  The differing coarse spectral 
characteristics of LBGs as described by \citet {aes03} prompted
slightly different identification expectations for each population.  
Spectra exhibiting Lyman $\alpha$ in emission and appropriate
continuum breaks were additionally inspected for the presence of 
interstellar absorption lines (usually weak) and a flat continua 
long-ward of Lyman $\alpha$ to aide in their identification.  
Spectra with Lyman $\alpha$ in absorption were inspected for the 
appropriate continuum break characteristics and multiple strong 
absorption lines with a somewhat reddened continua long-ward of 
Lyman $\alpha$.  At no time were the spectra required to have all of 
these features, although nearly all identified galaxies fulfilled 
these expectations.  

In practice, and depending on the wavelength coverage of an individual 
spectrum, usually 6 or more absorption lines were used to secure the 
redshift of a particular LBG. Due to the overall low signal-to-noise 
nature of these spectra, it becomes necessary to assign secure and 
probable LBG redshifts.  Both the secure and the probable LBGs exhibit 
the characteristic continuum profile.  Relevant expected stellar and 
interstellar metal lines used for identification included: Si II 1193, 
1260, 1304, 1527, Si IV 1394, 1403, C II 1335, C IV 1548, 1551, O I
1302, Fe II 1608, He II 1640, Al II 1671, Al III 1855, 1863 , and
CIII] 1909 among others.  Low redshift galaxies were identified by
comparison to templates of E+SO, E+A, and starbursting galaxies.
Since the break in the continua of low redshift galaxies near
4000\AA~can mimic the Lyman break of LBGs, low signal to noise
absorption spectra must be carefully checked to avoid
misidentification.  Emission lines, such as [O II] 3728, [O III] 4959, 
5007, [N II] 6584 and [S II] 6724 and typical absorption and/or
emission features such as the Hydrogen series, Ca II H \& K lines, Mg
II 2796, 2804, Mg I 2852, 5175, He I 4471, 5876, 6678, 7065 and Na I
5890, 5896 were used to confirm low redshift galaxies and Galactic stars.
%  make full photometric table(s) to submit electronically **
%  flag 0-6 categories?

In total, we acquired spectra of 529 objects from the 796 LBG
candidates that met either $u'$BR(I), $u'$VRI, or $u'$BVRI
photometric selection criteria.  Using an in-house analysis routine,
we interactively examined each one-dimensional and two-dimensional
spectrum individually for identifiable features following the
prescription described above.  All objects were then placed in one 
of the following seven categories: (0) stars, (1) secure $z<2$ galaxies, 
(2) probable $z<2$ objects, (3) probable $z>2$ objects, (4) secure
$z>2$ galaxies, (5) high redshift QSOs, and (6) unidentified objects.  
Objects placed in categories 2 and 3 were identified as low or high 
redshift objects respectively, but did not satisfy the identification
requirements as rigidly as categories 1 and 4.  The probable 
identifications usually resulted from lower signal-to-noise and lack
of apparent emission lines yet still showed reasonable continuum
breaks and the tentative identification of several to many absorption 
lines.  Some objects were placed in the probable LBG category due to a 
slight ambiguity in redshift. Objects were classified as
unidentified (category 6) if they displayed very low signal-to-noise 
(possibly from a portion of these objects falling outside the slitlet) 
or a strong ambiguity in redshift assignment.  These are listed for 
each field in Table~\ref{IDs} with the separate categories indicated 
by the parentheses. 

\begin{figure}
\begin{center}
\scalebox{0.6}[0.35]{\rotatebox{-90}{\includegraphics{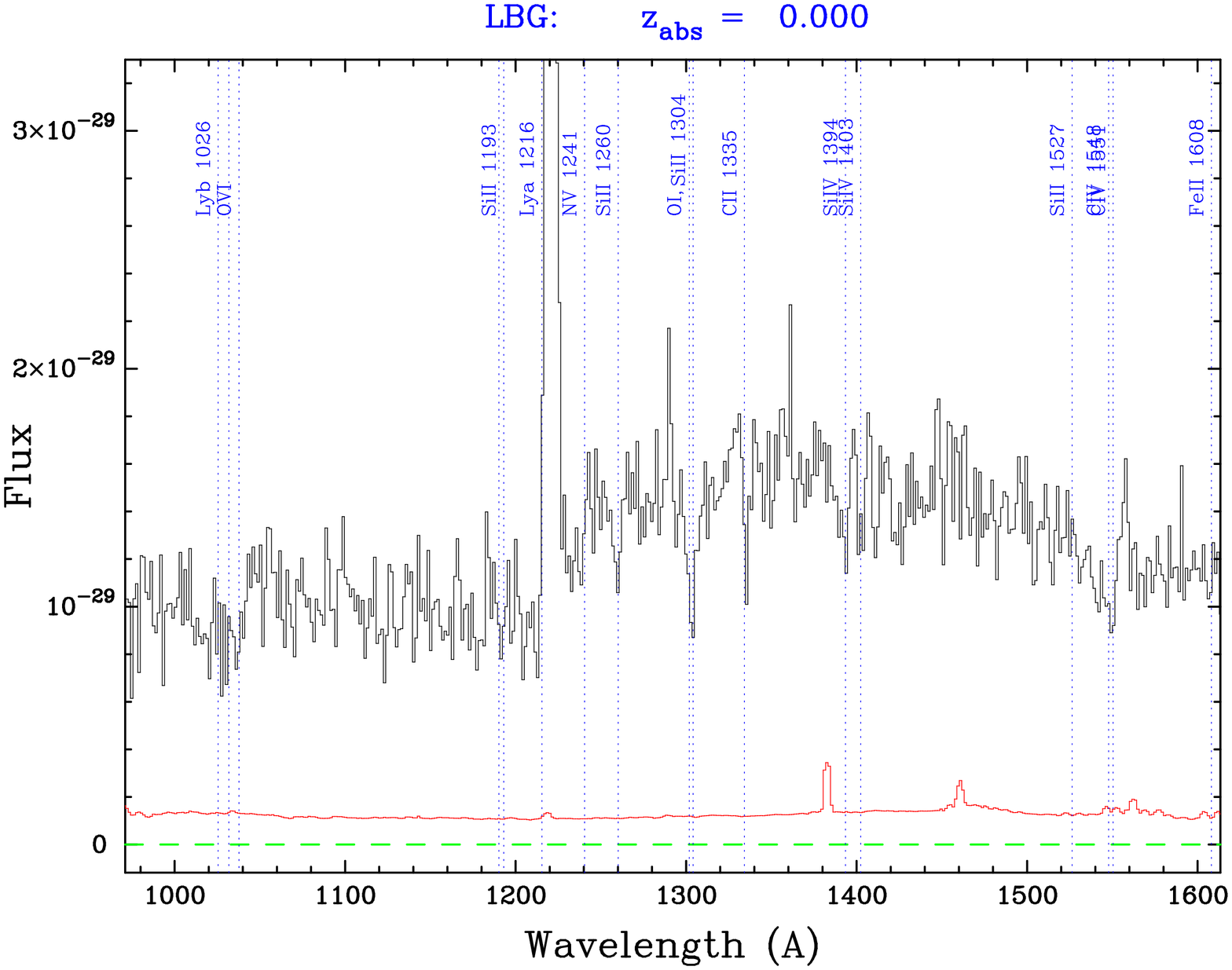}}}
\scalebox{0.6}[0.35]{\rotatebox{-90}{\includegraphics{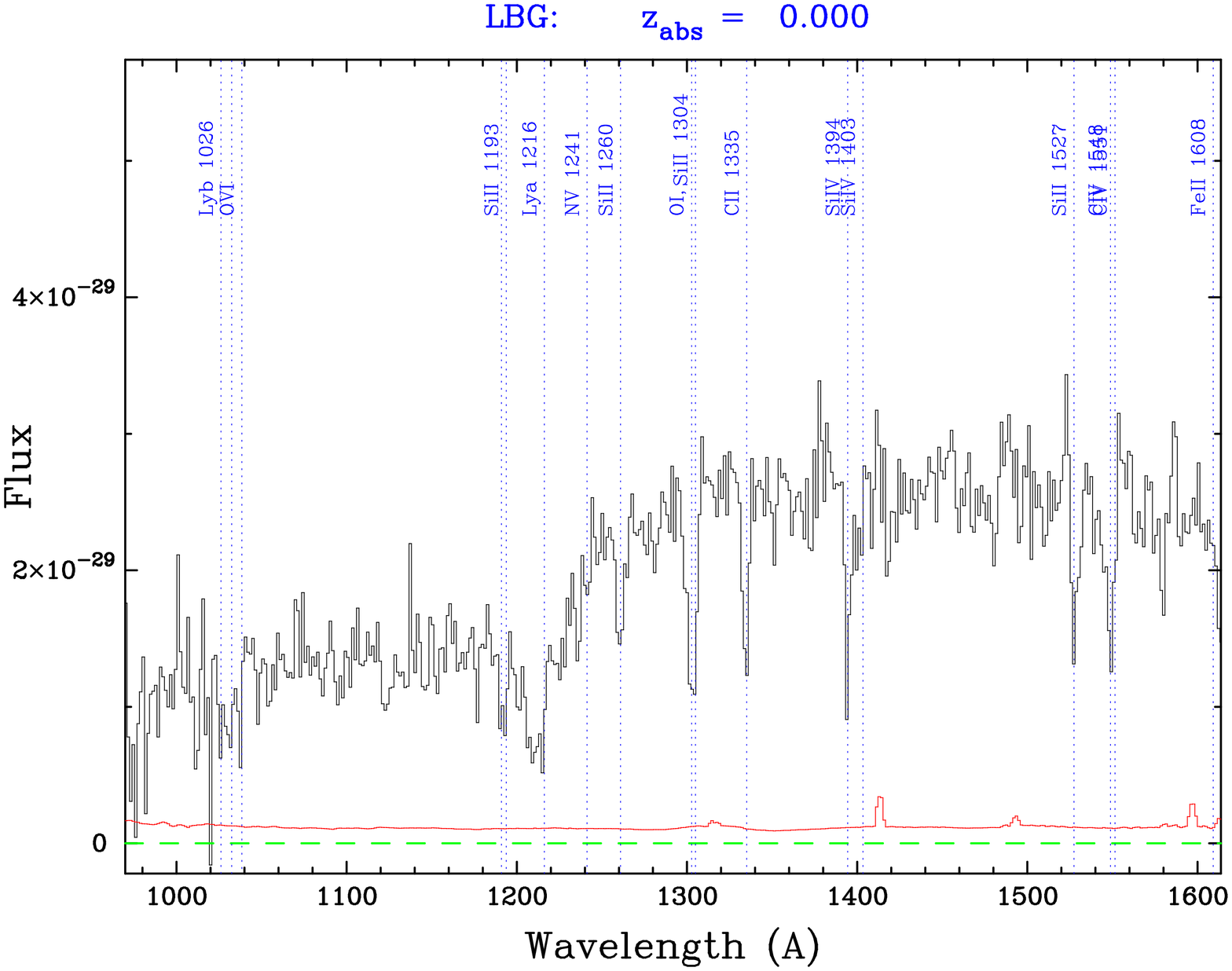}}}
\caption{\small Composite LBG spectra shifted to the rest-frame.  Each 
spectrum consists of multiple $z > 2$ galaxy spectra that exhibit similar
properties.  Expected interstellar absorption and emission lines are
labeled and indicated by the dotted lines.  Top: Composite spectrum of
25 LBG galaxies that display prominent Lyman $\alpha$ in emission.  
Interstellar absorption lines are typically weaker in this population.
A detectable break in their continua exists short-ward of 1215\AA. Bottom: 
Composite of 15 LBG spectra displaying Lyman $\alpha$ in absorption. 
These objects display stronger interstellar absorption lines and a
slightly reddened continuum.  Fluxing beyond rest-frame 1500\AA~in
each composite spectrum is less reliable for reasons discussed in the 
text. \label{composites}}
\end{center}
\end{figure}

Two composite LBG spectra are shown in Figure~\ref{composites}.  Each
spectrum was constructed by stacking the rest-frame spectra of
$\sim20$ individual LBGs that exhibit similar properties.  The upper 
spectrum is a composite of LBG spectra with prominent Lyman $\alpha$ 
emission.  This spectrum displays visibly weak (narrow) interstellar 
metal absorption lines and a relatively flat continuum long-ward of
1215\AA.  The Lyman $\alpha$ feature is redshifted with respect to the 
interstellar lines by an average of $\sim600$ km sec$^{-1}$, in
agreement with \citet {a03}.  This is interpreted to be due velocity
differences within the galaxy arising from stellar and
supernovae-driven winds.  The higher redshift of the Lyman $\alpha$ 
feature results from dust absorption of the resonantly scattered 
photons in the blueshifted portion of the expanding galactic shell.  
Emission-identified LBGs comprise $\sim46\%$ of our sample.  This is a 
greater fraction of LBGs of an equivalent type than found by \citet{s03} 
and may be due to a more conservative redshift identification 
process. The lower spectrum combines LBG spectra having Lyman $\alpha$ 
in absorption, revealing a somewhat reddened continuum and noticeably 
stronger (broader) interstellar absorption lines.  The assessment of
the overall continuum profiles of LBGs from our data is in agreement 
with the results of the analysis of intrinsic LBG properties by \citet 
{aes01,aes03}.  The fluxing of the spectra long-ward of
$\sim$1500\AA~in Figure 7 is not reliable due to the fact that no
order blocking filter was used for these observations.  Since standard 
stars used for fluxing tend to be very blue, their second order light 
can significantly affect the spectrum at observed wavelengths
long-ward of $\sim$6400\AA.  This corresponds to rest-frame
wavelengths long-ward of $\sim$1600\AA~for objects at $z\sim3$.
Moreover, the composite spectra comprise LBGs of varying redshifts.  
As a result, the effects from the flux degradation is spread over 
rest-frame wavelengths greater than $\sim$1500\AA.

\begin{figure}
\begin{center}
\scalebox{0.5}[0.46]{\includegraphics{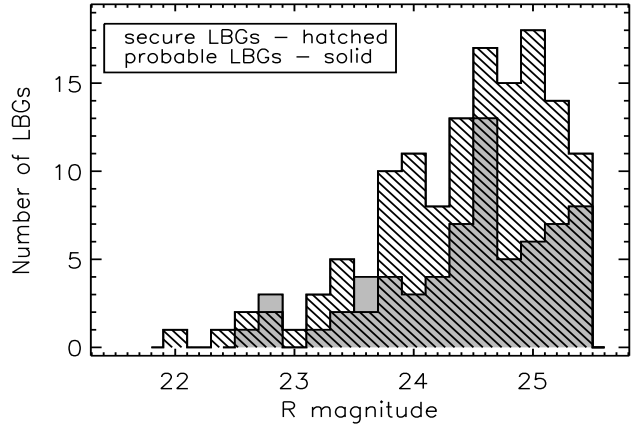}}
\caption{\small Histogram of the R magnitude distribution of the 
spectroscopically identified objects with $z > 2$.  The secure LBGs 
(hatched region) are shown with probable LBGs (solid region) overlayed.  
\label{Rmag}}
\end{center}
\end{figure}

\begin{figure}
\begin{center}
\scalebox{0.5}[0.46]{\includegraphics{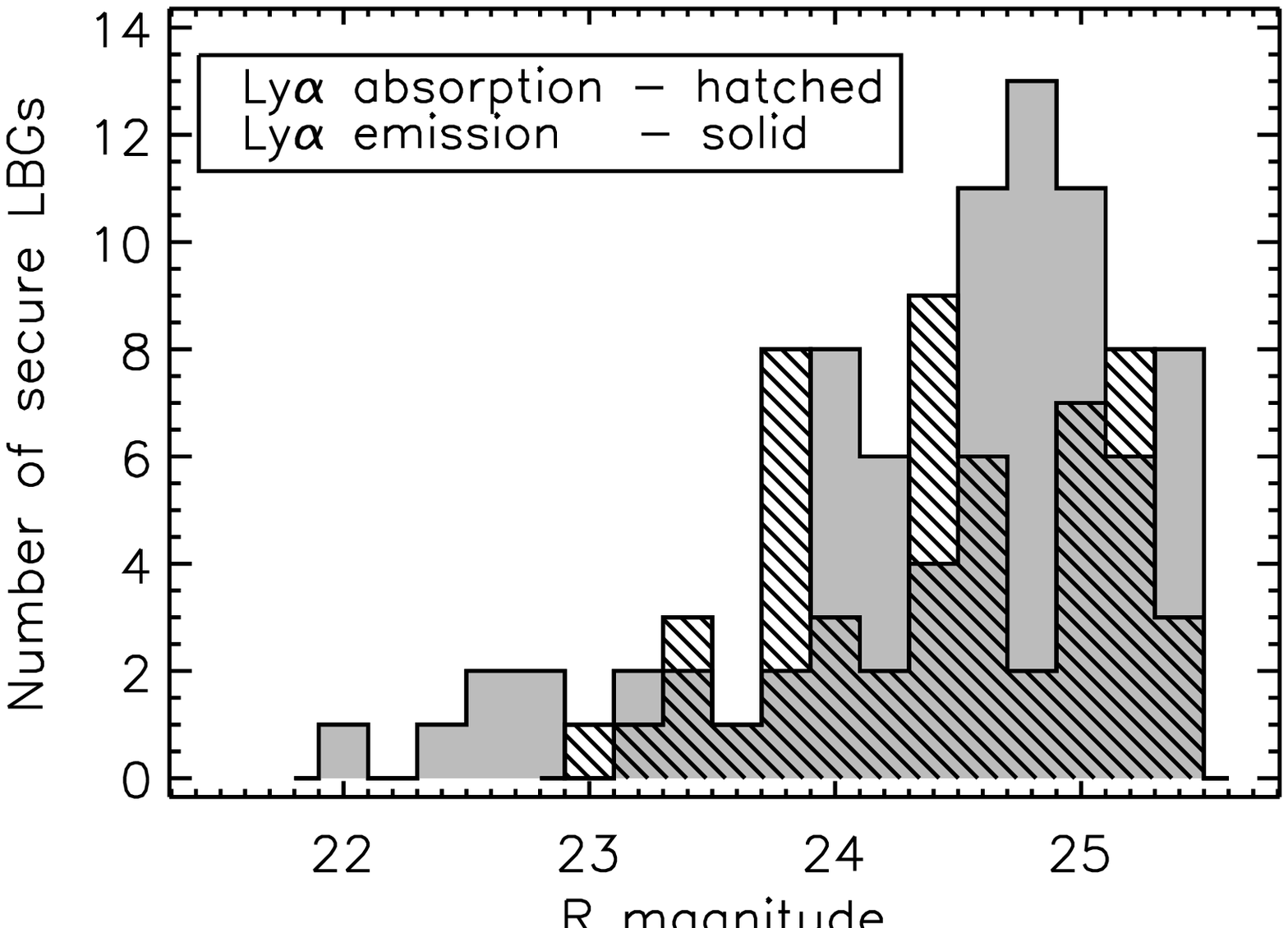}}
\scalebox{0.5}[0.46]{\includegraphics{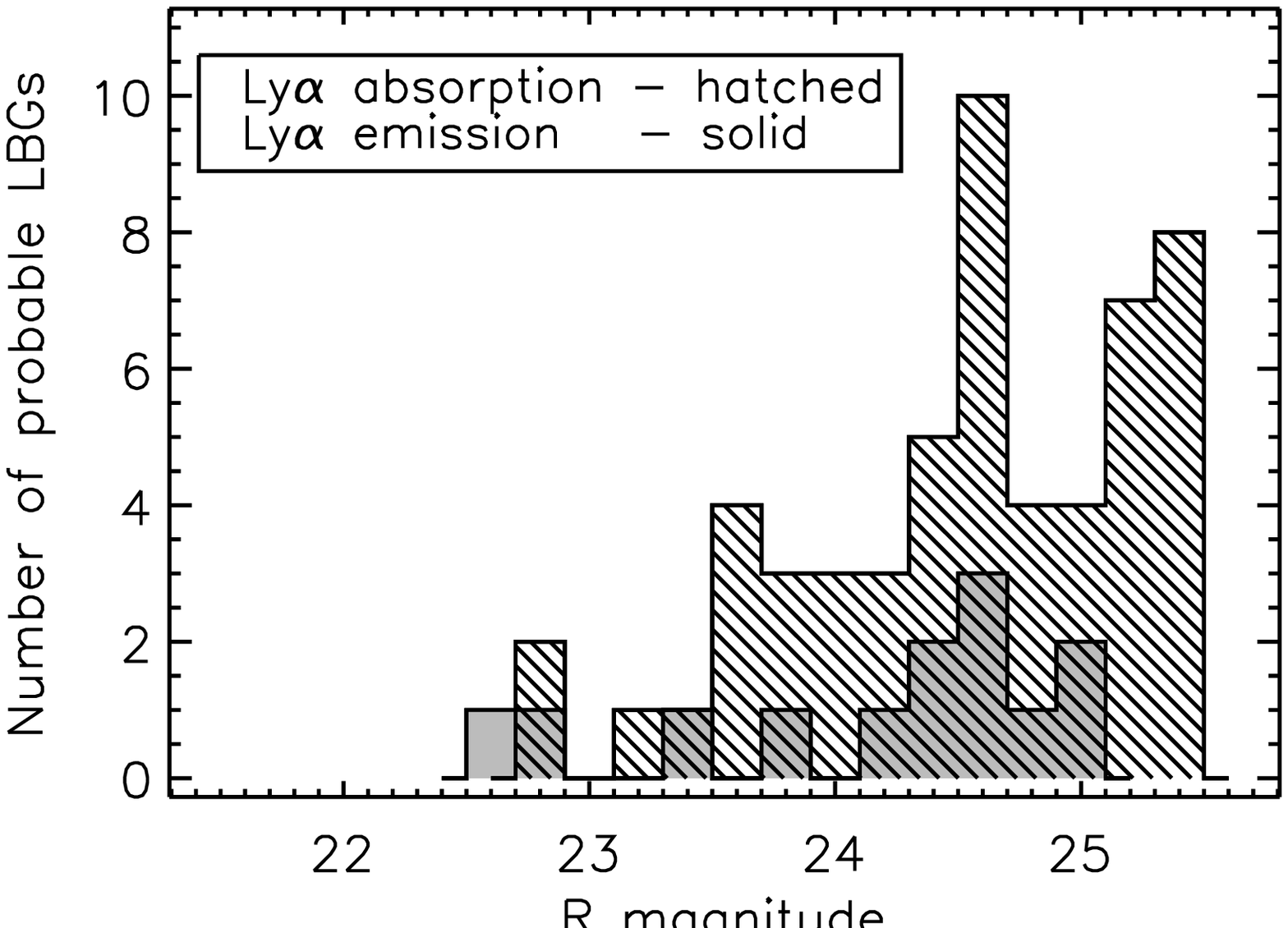}}
\scalebox{0.5}[0.46]{\includegraphics{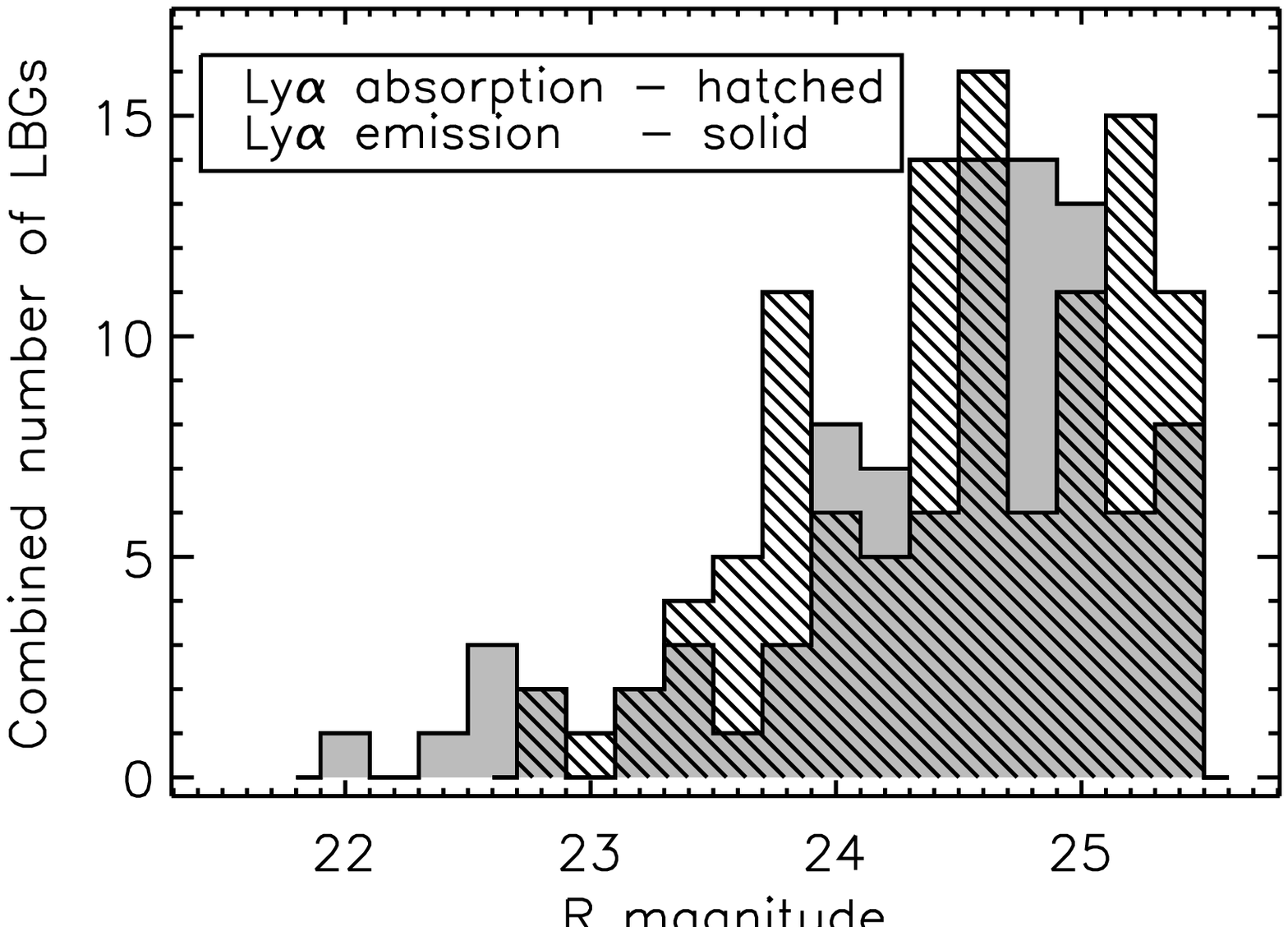}}
\caption{\small Histograms of the R magnitude distributions of the 
emission-identified (solid region) and absorption-identified (hatched
region) LBGs.  The upper and central histograms plot the secure and 
probable LBG distributions respectively.  The lower histogram plots the
distributions of the combined set of LBGs. \label{ea-Rmag}}
\end{center}
\end{figure}

To test our identification process for any bias toward magnitude, we 
performed a two-sided Kolmogorov-Smirnov (K-S) test of the secure and 
probable LBG R magnitude distributions. This resulted in an 88\% K-S 
confidence level for the null hypothesis, making it very unlikely that 
they are drawn from separate distributions.  Figure~\ref{Rmag} shows a 
histogram of the secure and probable LBGs versus R magnitude binned in 
$\Delta$R=0.2. Additionally, we tested for any bias in magnitude
toward LBGs displaying Lyman $\alpha$ in emission versus LBGs
displaying Lyman $\alpha$ in absorption since the presence of a 
strong emission feature could enhance the confidence of
identification.  This is especially the case for low signal-to-noise 
spectra.  Figure~\ref{ea-Rmag} presents R magnitude histograms of 
emission-identified LBGs with absorption-identified LBGs.  The K-S
tests on the secure, probable, and combined LBG distributions resulted 
in a 29\%, 46\%, and 51\% confidence for the null hypothesis,
respectively.  In all three cases, the values are consistent with the 
hypothesis that the secure emission-identified and
absorption-identified LBGs are drawn from the same parent distributions.  

%Performing the K-S test on the
%probable LBG distributions resulted in a 98\% confidence that the
%emission-identified and absorption-identified LBGs are likely to be 
%drawn from separate distributions.  This assessment is considered in 
%future analysis. However, a K-S test relies heavily on the median of 
%the distributions and may interpret 

The ability to securely identify an emission-identified LBG 
relative to absorption-identified LBG for the faintest objects tends 
to diminish with magnitude but will actually increase beyond R$\sim25$.  
This is reflected in Figure~\ref{ea-Rmag} where it can be seen that 
there are no emission-identified LBGs and 15 absorption-identified
LBGs with  R $>25$.  Although this is partially due to the presence of 
the Lyman $\alpha$ feature adding confidence to these lowest
signal-to-noise spectra, it is also an effect of the photometric
selection criteria which is designed to select galaxies based on the 
decrements in their continua.  Due to the finite magnitude depths of
the $u'$ and B-band images combined with strict $(V-R)$ and $(V-I)$ 
constraints, only the bluest objects selected with R $\sim 25$ and 
greater can meet the selection criteria.  The bluest LBGs are
typically emission-identified LBGs.  Because of this, the 15 R$>25$ 
absorption-identified probable LBGs are the least confident of the 
sample.

%       KS tests: 3&4 - 0.88   prob. same pop.
%       KS tests:  45 - 0.29   prob. same pop.
%                   3 - 0.46   prob. same pop.
%                 345 - 0.51   prob. same pop.

\begin{figure}
\begin{center}
\scalebox{0.5}[0.46]{\includegraphics{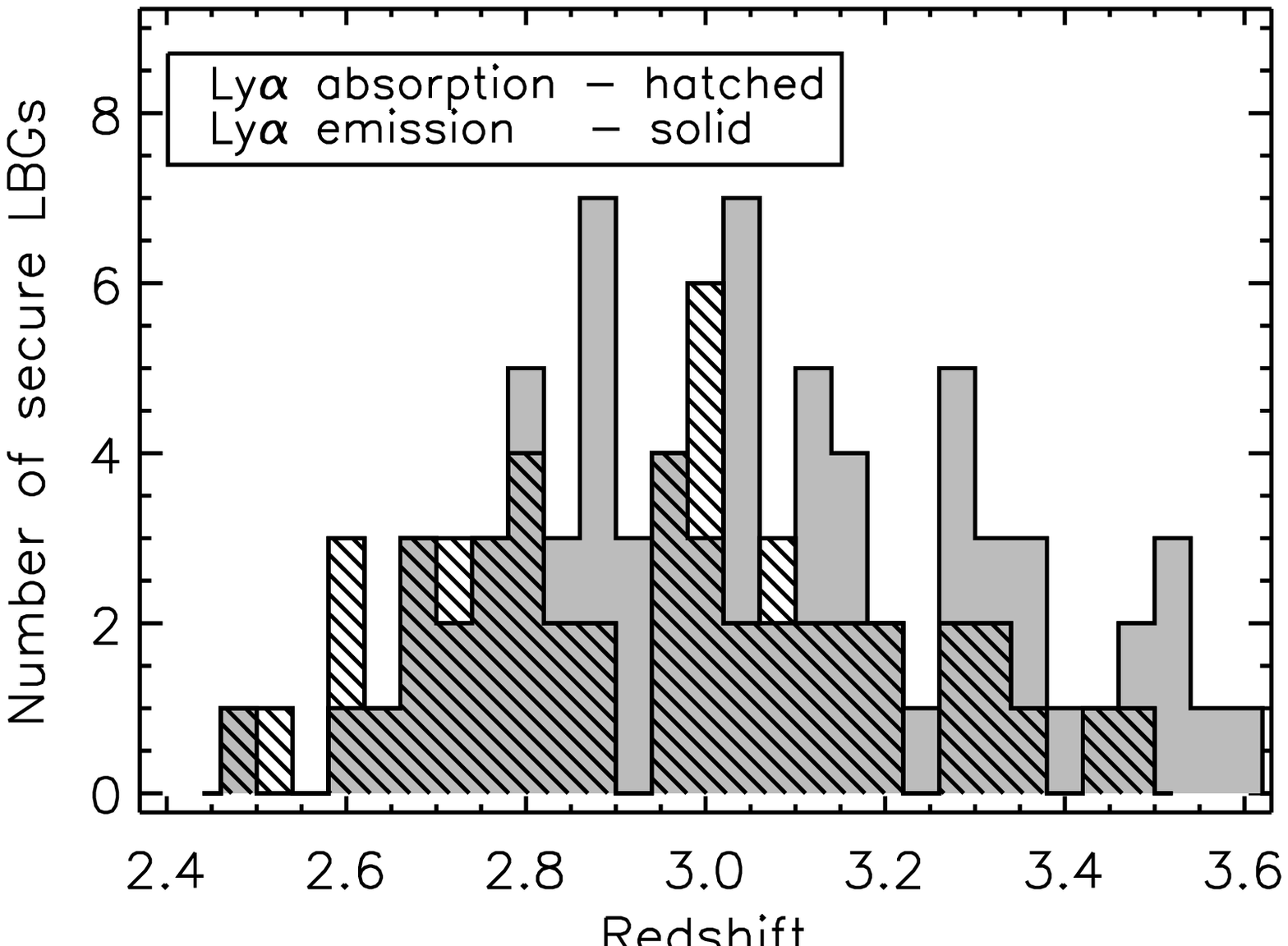}}
\scalebox{0.5}[0.46]{\includegraphics{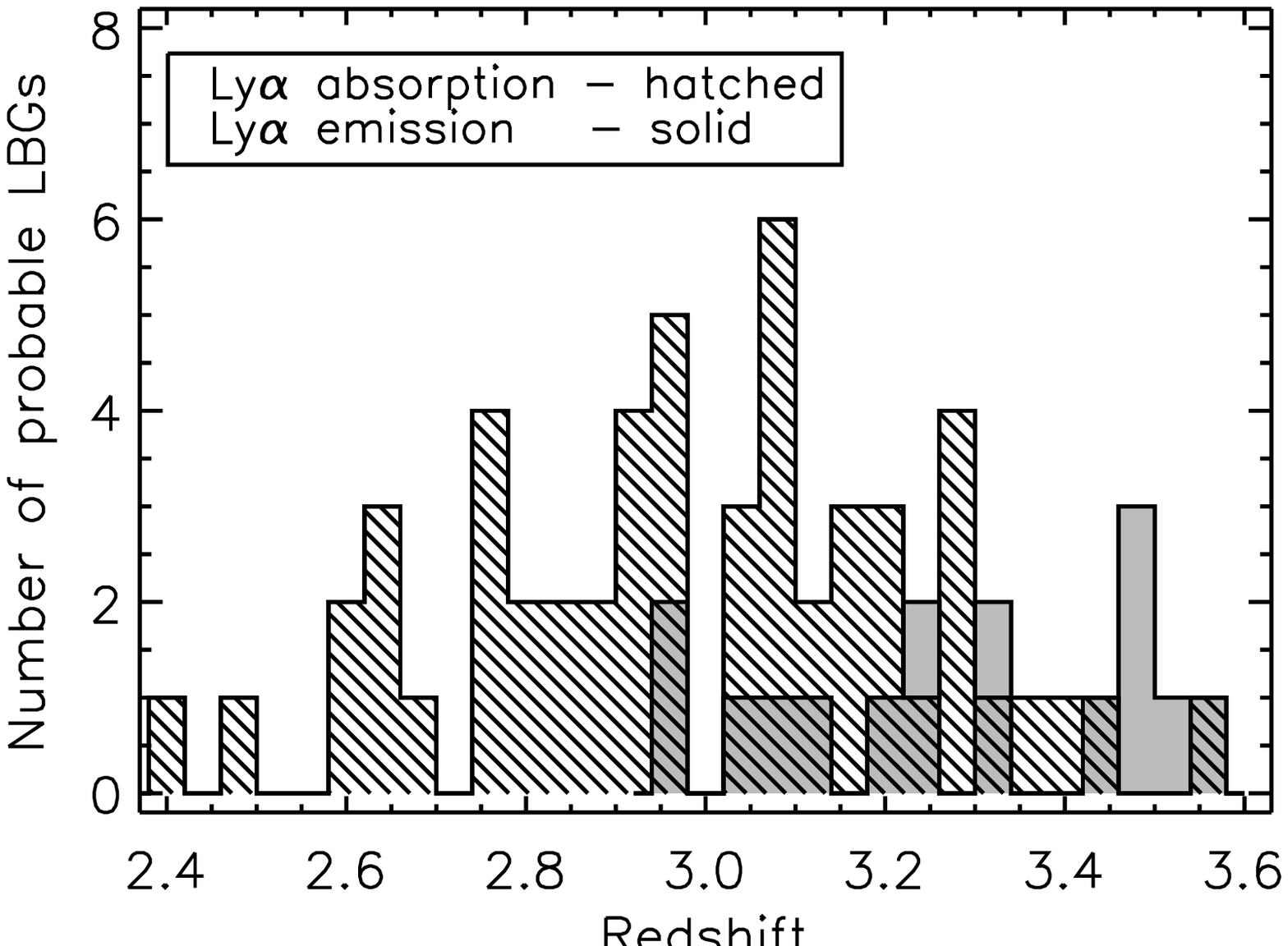}}
\scalebox{0.5}[0.46]{\includegraphics{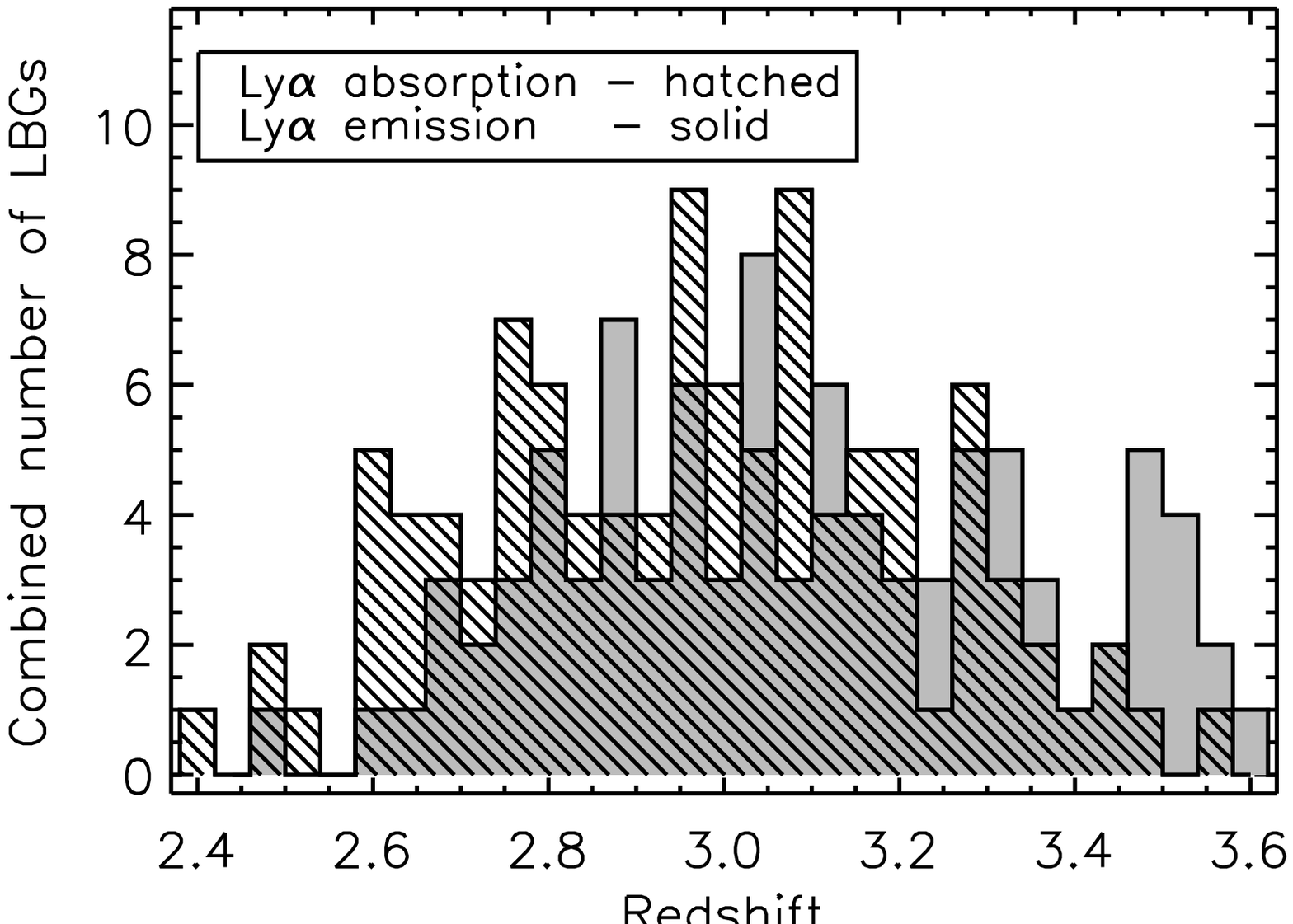}}
\caption{\small Redshift distributions of LBGs displaying Lyman
$\alpha$ in absorption (hatched region) and emission (solid region).
The secure LBG identifications are presented in the upper histogram 
and probable LBG identifications are presented in the center histogram. 
The lower histogram displays both secure and probable LBGs combined.  
The broad similarities in the absorption-identified redshift 
distributions reinforces the overall probable redshift assignments.  
\label{ea-z}}
\end{center}
\end{figure}

The redshift distribution of the identified LBGs is another regime to
test our identification process and lends insight into the overall 
distribution of LBGs detected by this selection method. 
Figure~\ref{ea-z} presents histograms that display secure, probable, 
and a combination of the two LBG categories in the upper, central, and
lower histograms respectively.  The broad agreement in the 
absorption-identified LBG redshift distributions lends confidence to
the overall redshift assignments.   The separate sets of 
emission-identified and absorption-identified LBGs show a small offset 
in the centers of their distributions.  The secure emission-identified 
LBG distribution centers at $z=3.07 \pm0.33$ while the secure 
absorption-identified LBG distribution centers at $z=2.92 \pm0.25$.  
Nevertheless, a K-S test on the two secure distributions resulted in a 
21\% confidence of the null hypothesis and is consistent with the two 
distributions being drawn from a similar parent distribution.  

Performing the K-S test on the probable LBG redshift distributions 
resulted in a 99\% confidence that the emission-identified and
absorption-identified LBGs are likely to be drawn from different
distributions.  This assessment is considered in future analysis and 
can be seen by the apparent lack of probable emission-identified LBGs 
with $z<2.9$. 
%  ** KS test:   45 - 0.21  prob. same pop.
%                 3 - 0.004 prob. same pop.
%               345 - 0.04  prob. same pop.
%Two reasons for this may be due to specific emission features that 
%help identify $z<2$ galaxies.  These 
%interlopers are often identified in emission by their [O II] 3727 and 
%[O III] 4959, 5007 transitions.  Lyman $\alpha$ cannot be mistaken for 
%an [O III] transition at redshifts less than $z\sim3.1$ and therefore
%may add to the confidence of these redshift assignments. Additionally,  
%objects greater than $z\sim3.0$ with preliminary [O II] 3727 emission
%line identifications can find difficulty in a secondary confirmation 
%using possible [O III] 4959, 5007 transitions since these transitions 
%can be either redshifted to a region where there are many
%competing bright sky emission lines that $LRIS$ is less capable of 
%extracting or may be redshifted out of the wavelength range of the
%particular spectrum.  These factors contribute to the uncertainty in 
%the redshift assignment of these emission-identified galaxies.
Although emission features add to the general confidence of spectral
identifications, emission-identified Lyman break objects at the higher 
end of the redshift range probed by our method become less confident
since the majority of their expected stellar and interstellar features
begin to compete with a dense array of night sky emission lines.  This
is also the case for the most common low redshift interlopers. The
LRIS instrument is less capable of subtracting these lines effectively
for very low signal-to-noise spectra.

From the full spectroscopic sample, we arrive at a secure (category 4) 
LBG number density of 0.37 arcmin$^{-2}$ with a field-to-field scatter 
of $\pm0.21$ arcmin$^{-2}$ and $0.56\pm0.27$ arcmin$^{-2}$ for the 
combined set of secure and probable LBGs (categories 3 and 4).  
Approximately two-thirds of the photometrically selected R$<25.5$ LBG 
candidates have spectroscopic coverage and approximately $63\%$ of the 
obtained spectra are identified.  Assuming similar results for the 
remaining R$<25.5$ LBG color candidates gives a projected LBG number 
density of $0.78\pm0.46$ arcmin$^{-2}$ for the secure LBGs and 
$1.21\pm0.49$ arcmin$^{-2}$ for the secure and probable LBGs.

%------------------------------------------------------------------------

\subsection{The Photometric Selection Function}

\begin{figure}
\begin{center}
\scalebox{0.45}[0.41]{\includegraphics{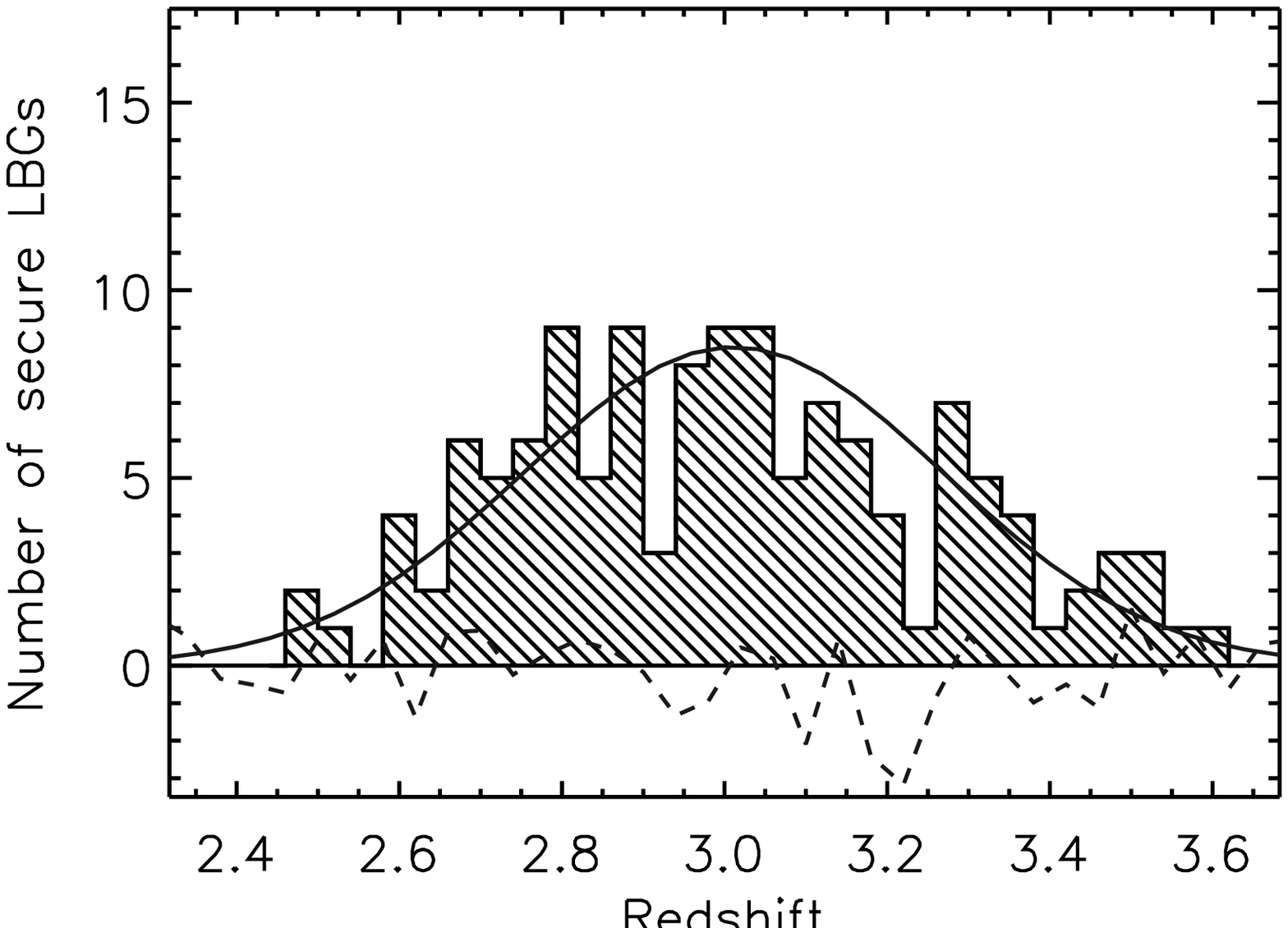}}
\scalebox{0.45}[0.41]{\includegraphics{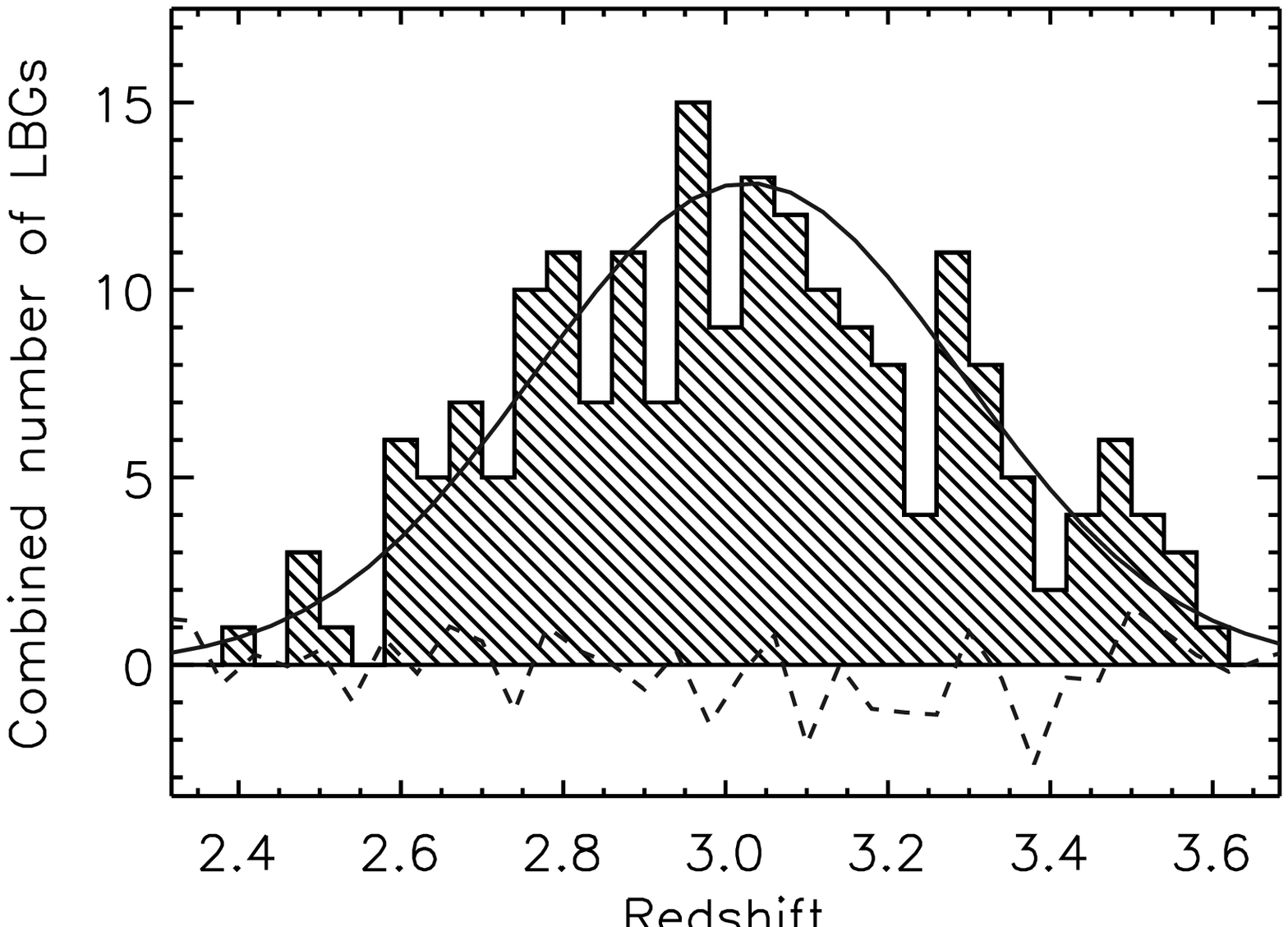}}
\scalebox{0.45}[0.41]{\includegraphics{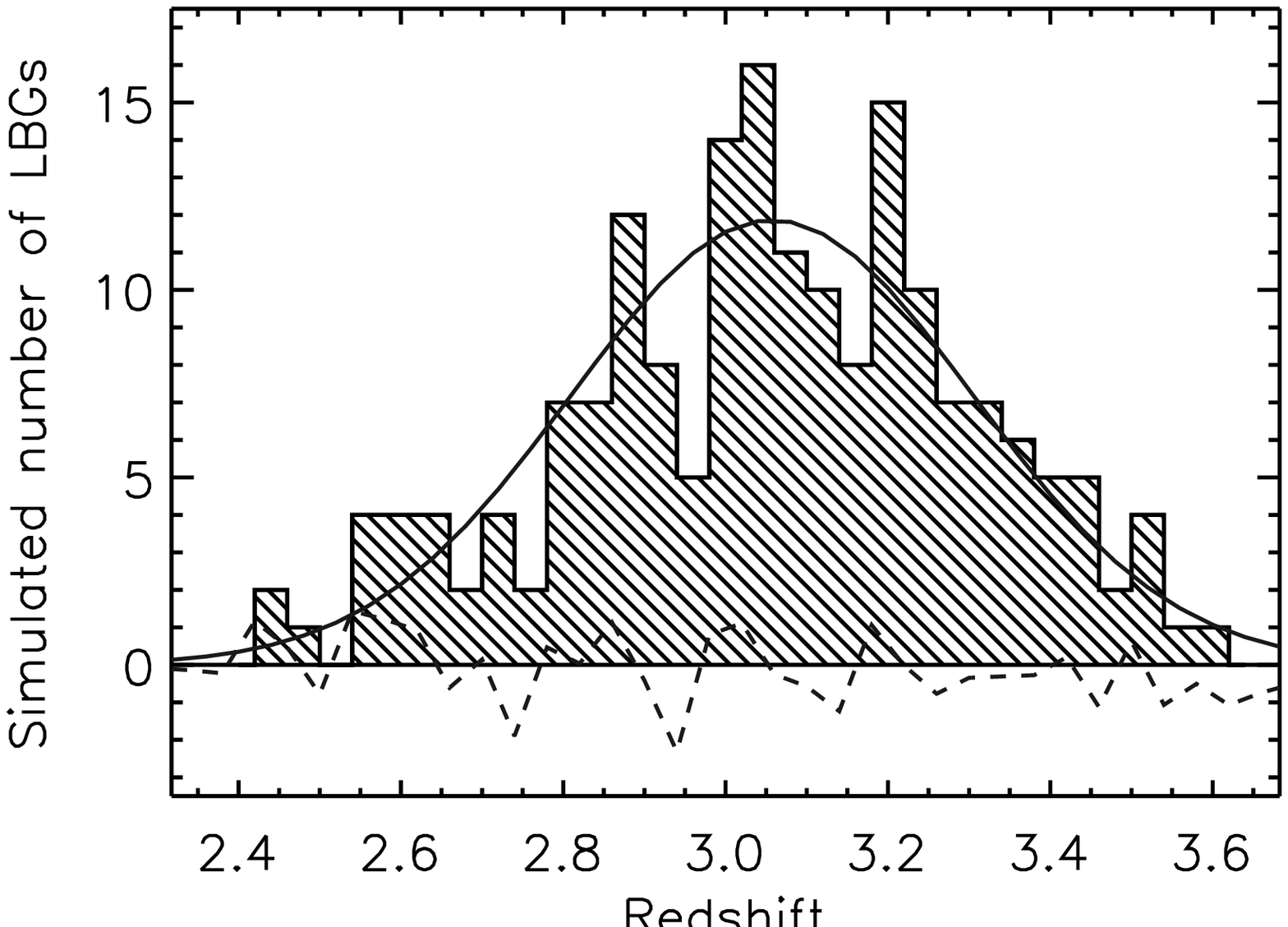}}
\caption{\small Real and simulated redshift histograms.  The top histogram 
represents the redshift distribution of 137 LBGs from the data with secure 
spectroscopic redshift identifications. The central histogram includes an 
additional 74 objects with probable redshift identifications.  The LBG 
selection function is depicted by the solid curve in both cases.  The lower 
histogram displays the results of a simulation of the observed redshift
distribution accounting for the characteristics of the instrument,
filters, and observed photometric uncertainties.  The simulation
assumes a flat background distribution of LBGs over the redshift path 
$2.4<z<3.6$. The normalized residuals to the Gaussian function fits
with respect to the expected Poisson fluctuations per bin are
indicated by dashed curves.  
\label{selectfunc}}
\end{center}
\end{figure}

The selection criteria described in \S~3 were designed to be sensitive 
to objects over a redshift path from $2.6<z<3.4$.  The previous
large LBG survey at $z\sim3$ by \citet {s03} probes a similar redshift
path with a distribution of $\langle z\rangle=2.96\pm0.29$ for their 
catalog of 940 identified spectra.  They used their spectra to
establish several fundamental properties of LBGs at $z\sim3$.  Our 
color selection technique was intentionally designed to select LBGs 
over the same redshift path to utilize the LBG properties defined
by Steidel and coworkers. The redshift distributions of the LBGs 
detected by this survey are presented in the upper two histograms of 
Figure~\ref{selectfunc}. The top histogram displays the 137 secure LBG 
redshifts from $2.4<z<3.6$ binned in $\Delta z=0.04$.  The central 
histogram represents the distribution of the combined set of LBGs, 
which includes 74 additional probable LBG redshifts.  The secure 
LBG distribution has a mean of $\langle z\rangle=3.00$ with a standard 
deviation of $\pm0.32$.  The combined set of 211 objects is nearly 
identical with $\langle z\rangle=3.02\pm0.32$.  These distributions 
include the five faint $z\sim3$ QSOs discovered in this survey. 

Observed redshift distributions combine the actual LBG 
redshift distribution with the effects of the photometric selection 
method.  To measure the clustering properties of LBGs associated with 
DLAs, the observed LBG densities per redshift interval must be
corrected for the photometric selection effects caused by the 
efficiencies and transmissions of the instruments and filters as well 
as the photometric uncertainties.  These corrections can be used to 
recover the true LBG background redshift distribution from the
observed LBG redshift distributions once the true photometric
selection function is determined.  Both observed LBG redshift 
distributions from the data (Figure~\ref{selectfunc}) were fit by 
Gaussian functions (solid curves) using the mean, standard deviation, 
and the same total number of objects of each distribution.  The 
normalized residuals to the fits with respect to the expected Poisson 
fluctuations per bin are $-0.05\pm0.95$ ($0.03\pm0.89$) for the
secure (combined) set of LBGs and are over-plotted on 
Figure~\ref{selectfunc} as dashed curves.  The fits indicate that in 
both cases the product of the photometric selection function and the 
true LBG redshift distribution are well described by Gaussian functions. 
%  In the case of this survey, the LBG background redshift 
%  distribution is the average redshift distribution over nine fields.

To model the true photometric selection function, we generated a catalog 
of 1000 starbursting galaxies with random magnitudes drawn uniformly 
from $23.5<R<25.5$ and random redshifts from $2.4<z<3.6$.  These 
galaxies were then assigned respective $u'$BVI magnitudes according to 
their assigned redshifts using the theoretical galaxy templates of
\citet{y00}.  The quantum efficiency of the CCDs, atmospheric
attenuation, and transmission characteristics were factored into the 
response of each filter.  We fit a function to the photometric errors
of the data with respect to magnitude and assigned Gaussian random 
errors multiplied by this function to each galaxy accordingly.  Under 
the assumption that the LBG spatial density varies slowly with 
redshift, we assumed a uniform background LBG redshift distribution
and corrected the expected observed LBG background to include 
the fact that a standard candle diminishes by $\sim1.1$ magnitudes 
from $z=2.4-3.6~(h=0.72, \Omega_M=0.3, \Omega_{\Lambda}=0.7)$.
In doing this, we incorporated the fact that our survey is magnitude 
limited by only considering objects with a corrected magnitude of
R$\le25.5$.  

The results of this simulation are shown in the lower histogram in 
Figure~\ref{selectfunc}.  This simulation shows that the product of a 
uniform LBG redshift distribution and the simulated true photometric 
selection function is well described by a Gaussian (solid line) and
results in a distribution with nearly identical mean and standard
deviation as the data. In an identical manner as the data, the
normalized residuals of the Gaussian fit to the simulation are 
over-plotted as the dashed curve and have a mean of $-0.11\pm0.76$.  

Since the simulated redshift distribution which incorporated the
effects of our photometric selection method and assumed a uniform
background LBG redshift distribution is shown to be described well 
by a Gaussian function, and since the observed redshift distributions 
can similarly be described by Gaussian functions, it is reasonable to 
assume a uniform background number density from $2.4<z<3.6$.
Knowledge of the LBG background number density per redshift interval 
is required for proper determination of the DLA-LBG clustering
amplitude.  Although the true background LBG number density from 
$2.4<z<3.6$ may not be strictly uniform, any slow varying
non-uniformity in the large-scale distribution of LBGs will not affect 
the DLA-LBG clustering amplitude significantly.  Indeed, the spatial 
distribution of LBGs is clustered on small scales, but bin-by-bin 
averaging over the 9 fields and 11 $z\sim3$ DLAs in this survey will 
in effect ``wash out'' an individual field's clustering signal.

\section{COLOR SELECTION EFFICIENCIES AND CRITERIA REFINEMENT}

The $u'$BVRI photometric color selection technique presented here is 
an efficient means to detect LBGs over the intended redshift path of
$2.6<z<3.4$.  Although the original intent of this survey was 
to perform $u'$BVRI photometric color selection in all 9 fields, 
various instrument and imaging concerns, described above, required the 
use of a subset of the full color criteria in four fields: 
LBQS0056+0125, PKS0336-017, PC1643+4631A, and JVAS2344+3433.  
Restricting these fields to a subset of the $u'$BVRI filters 
did not significantly hinder our ability to detect LBGs. For the five
fields with $u'$BVRI imaging, the number density of LBG candidates 
meeting $u'$BVRI color criteria allowed room on each slitmask to
include objects meeting lesser color constraints.  This allowed an 
additional opportunity to examine the efficiencies of these alternate 
criteria.

Table~\ref{eff} summarizes the efficiencies of the $u'$BVRI color
selection criteria and subset criteria from the spectroscopic data.
Adhering to the imposed  $u'$BVRI color selection criteria outlined in 
$\S$~3.1 and the spectroscopic identification process in $\S$~4.1, we 
find 79\% of the identified objects selected for spectroscopy as
secure and probable $z>2$ Lyman break objects, 0\% were Galactic
stars, and 21\% were secure or probable $z<2$ galaxies.  The
efficiency is clearly influenced by the conservative nature of our 
redshift assignments and subjective object placement in the probable 
categories. Re-categorizing the probable spectroscopic placements 
would lead to as much as 95\% and as few as 59\% of the identified 
objects with $z>2$.

Nearly 60\% of the unidentified spectra can be attributed to
cirrus during the 2003 February observing run for the PKS0336-017
field and an instrument failure for the 2002 December observations
of the PSS1057+4555 and JVAS2344+3433 fields in which the CCD shutter
could not be closed.  In the latter case, we corrected for the
additional light striking the CCD during read-out by subtracting a 
one-second image taken immediately after each observation image.  
This was surprisingly effective in removing global flux levels, but 
was by no means perfect, especially in the complete removal of bright 
galaxy and star trails.

\begin{figure}
\begin{center}
\scalebox{0.35}[0.38]{\rotatebox{90}{\includegraphics{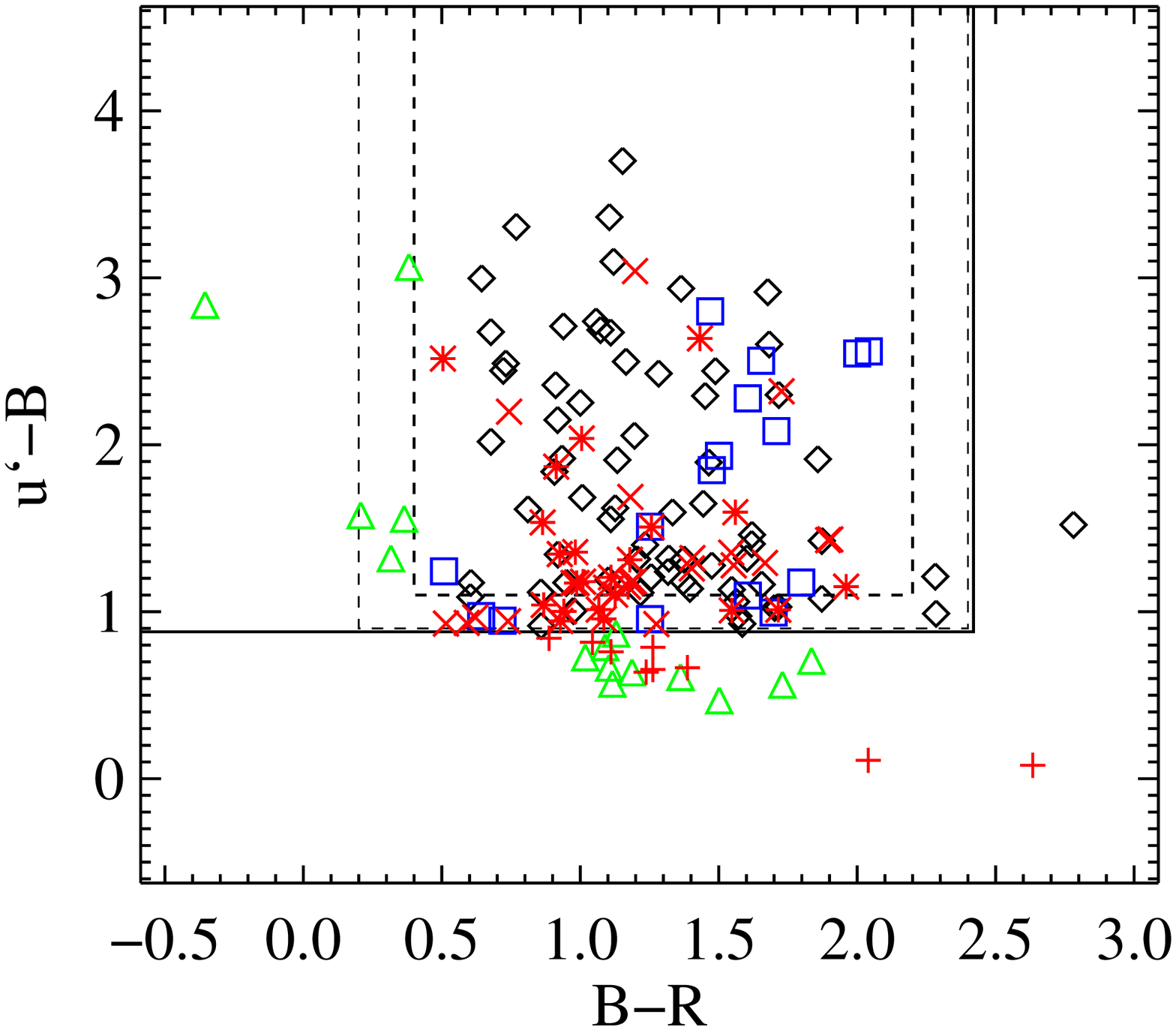}}}
\scalebox{0.35}[0.38]{\rotatebox{90}{\includegraphics{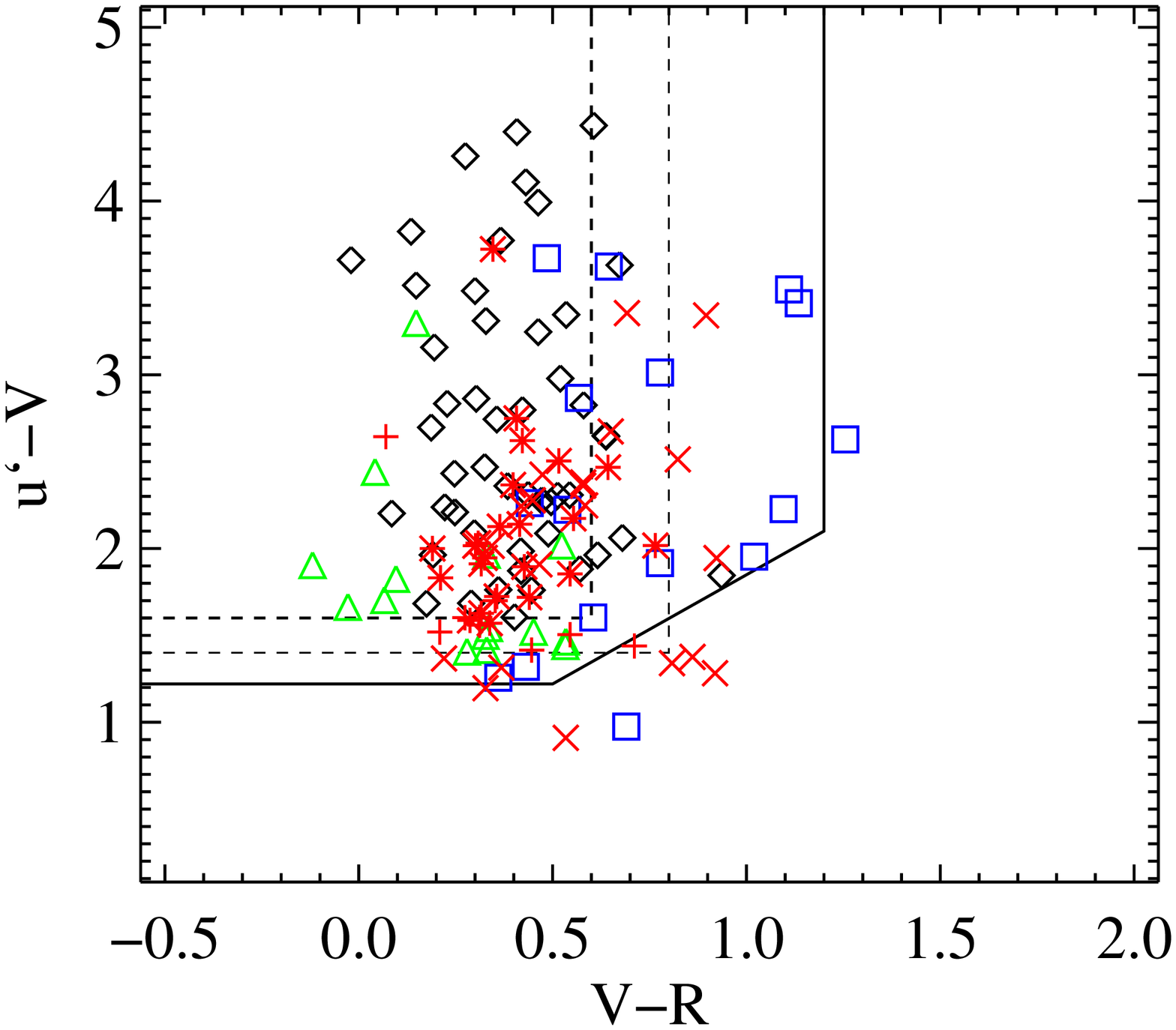}}}
\scalebox{0.35}[0.38]{\rotatebox{90}{\includegraphics{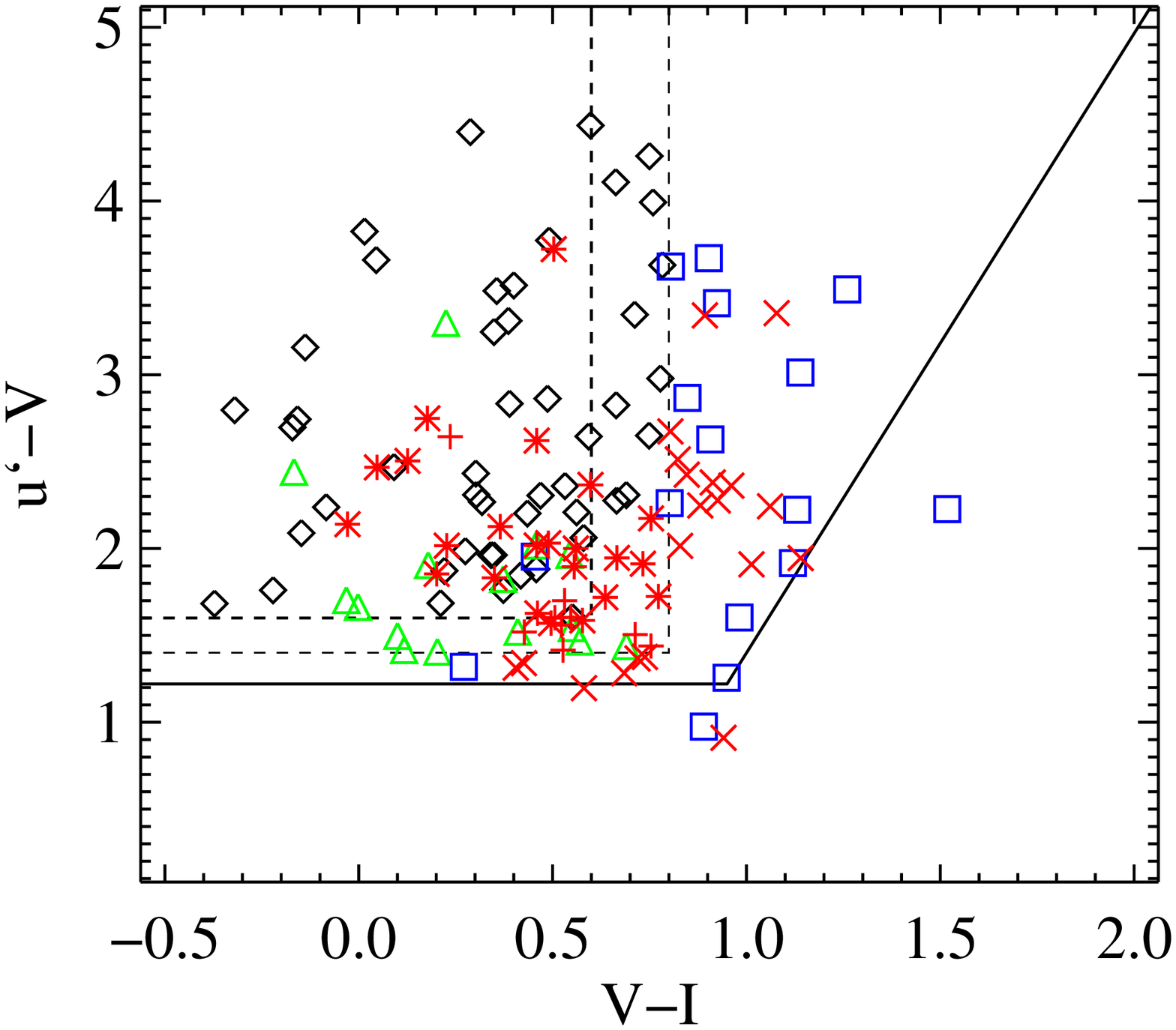}}}
\caption{\small Data from five fields with $u'$BVRI color selection
are plotted to re-assess the original photometric selection criteria 
(dashed lines).  The $z > 2$ LBGs meeting $u'$BVRI color criteria are 
marked with black diamonds.  Those meeting only $u'$BRI or $u'$VRI 
color criteria are marked with blue squares or green triangles 
respectively.  Interloping objects with $z < 2$ redshifts from the
data meeting $u'$BVRI color criteria are marked with red asterisks 
whereas those meeting only $u'$BRI or $u'$VRI color criteria are 
marked with red plus `+' signs or `X's respectively.  For the objects 
that are non-detections in the $u'$-band, the points represent the 
2$\sigma$ lower limit to their $(u'-B)$ and $(u'-V)$ colors.  
Spectroscopic analysis of confirmed LBGs shows that a fraction of 
the galaxy population at $z\sim3$ is missed using the full $u'$BVRI 
criteria.  To probe a more complete sample of galaxies at $z\sim3$ at 
the risk of lower efficiency, a possible refinement to the $u'$BVRI 
photometric selection is shown (solid lines) in the upper-left 
portions of the plots. \label{refine}}
\end{center}
\end{figure}

A review of the placement of spectroscopically confirmed LBGs on the
color-color plots allows an opportunity to assess and possibly refine 
the color selection criteria.  Figure~\ref{refine} displays the
spectroscopic results from the five fields (PSS0808+5215, PSS0957+3308,
BRI1013+0035, PSS1057+4555, and PSS1432+3940) that have 
$u'$BVRI color selection.  Zoomed-in views of the selection regions 
are shown and indicated by the boxed regions (dashed lines). LBGs 
(both secure and probable) selected by their $u'$BVRI color criteria 
are shown as black diamonds,
those selected by only their $u'$BR(I) color criteria are shown as blue
squares, and those selected by only their $u'$VRI colors are shown as 
green triangles.  All objects with $z < 2$ identifications are shown
in red.  The $z < 2$ galaxies and Galactic stars meeting $u'$BVRI
color criteria are shown collectively as asterisks and those meeting 
$u'$BR(I) or $u'$VRI color criteria are shown as plus `+' signs or
`X's respectively. 

The practice of obtaining spectra of the lower priority candidates 
revealed and helped quantify the fraction of the galactic population 
at $z\sim3$ that is missed when using the full set of $u'$BVRI
filters. Additionally, objects detected and confirmed using only 
$u'$BR(I) color criteria are missed when using only $u'$VRI color 
selection and vice-versa.  This can be seen in Figure~\ref{refine}.  
The blue squares to the right of the selection region in the $(u'-V)$
vs. $(V-R)$ plot were detected and confirmed using $u'$BR(I) color 
selection. Similarly, the green triangles below and to the left of the 
selection region in the $(u'-B)$ vs. $(B-R)$ were detected and
confirmed using $u'$VRI color selection.  The original $u'$BVRI color 
criteria (dashed lines) provide a very efficient method to discern the 
largest possible number of Lyman break objects from fields objects
with a minimum number of undesired low-redshift objects.  To acquire a 
more complete sample of all populations of galaxies at $z\sim3$, the 
data suggest refinements to the selection regions if either all or a 
subset of the $u'$BVRI filters are used.

From the $(u'-B)$ vs. $(B-R)$ plot in Figure~\ref{refine}, it appears
that the original $(u'-B)$ color cut is effective and any relaxation
of the criterion beyond 0.2 magnitudes would include a large number of 
less desired field objects (review Figure~\ref{dataplots} for the field 
object distributions).  Additionally, the extension of  the $(B-R)$ 
color cut to low values would help to include or confirm objects detected 
via $u'$VRI color selection.   Due to the lack of field objects in this 
region, this extension does not risk a loss to efficiency.   A refinement 
to this plot (solid lines) would be to set the $(u'-B)$ criterion to 
$(u'-B) > 0.9$ and modify the original color criterion from $0.6<(B-R)<2.1$ 
to $(B-R)<2.4$.  

A modification of the color criteria in the $(u'-V)$ vs. $(V-R)$ 
color-color plot (solid lines) is intended to include objects that 
would be detected via $u'$BR(I) color selection.  This refinement
would modify the current criteria from $(u'-V)>1.6, (V-R)<0.6$ to
$(u'-V)>1.2, (V-R)<1.2, (u'-V)\ge 1.4~(V-R)+0.5$.  Unfortunately, the 
distribution of low redshift interlopers extends into the modified 
selection region.   This refinement would include a broader sample of 
galaxies at the risk of diminishing the LBG detection efficiency.  

Several objects chosen by only their $u'$BR(I) colors (blue 
squares), show up to the right of the original selection region in the 
$(V-R)$ and $(V-I)$ color-color plot.  To include these objects and to 
search for others with reasonable SEDs without the high cost of many 
interfering field objects, a suggested refinement to the $(u'-V)$ vs. 
$(V-I)$ color criteria is shown (solid lines).  The suggested refinement
involves modifying the original color cuts from $(u'-V)>1.6, (V-I)<0.6$ 
to $(u'-V)>1.2$ and $(u'-V)\ge 3.8~(V-I)-2.6$.  

Finally, in our analysis of the 529 color selected objects, we also 
analyzed as many field objects as possible that randomly fell into our
slitlets.  These objects provide a small sample of the full
population of galaxies at high redshift free from color selection bias.  
From the 132 serendipitous field objects, and using the categories 
described above, we found 0 Galactic stars, 16 
secure $z<2$ galaxies, 27 probable $z<2$ objects, 8 probable $z>2$ 
galaxies, 4 secure $z>2$ galaxies, 0 QSOs, and 77 unidentified
objects.  Three of the secure $z>2$ galaxies are $z>3.65$ and would
not have been detected by our selection criteria.  The fourth galaxy 
is an emission-identified LBG at $z=3.263$ that falls below the 
spectroscopic magnitude cut-off for our survey of R=25.5.  Of the 
probable $z>2$ galaxies, 3 have $z>3.55$ and the remaining 5 were 
found to have $2.94<z<3.29$ and R$>25.5$. Although the 55 serendipitous 
objects identified sample a very small area of the sky relative to the 
entire survey, none of these objects were found within the redshift 
path probed by our selection method and brighter than our magnitude 
limit of R=25.5.  These statistics are consistent with the idea that 
our selection method detects the majority of objects at $z\sim3$.

%---------------------------------------------------------------------

\section{SUMMARY}

This paper presents the first spectroscopic survey designed to detect 
galaxies associated with known DLAs at $z\sim3$.  A primary goal of
this survey is to measure the spatial clustering of LBGs and DLAs to
determine the DLA-LBG cross-correlation and infer the typical mass of 
DLAs in the context of CDM cosmology.  Since DLAs may have a range of
mass, a survey for LBGs in multiple QSO fields with known DLAs is 
required to achieve proper statistics.  We present and analyze the 
imaging and spectroscopic data for objects in 9 QSO fields having 11 
known $z\sim3$ DLAs, including one DLA discovered by this survey, for 
the purpose of measuring the DLA-LBG cross-correlation.  

We selected a total of 796 objects as LBG candidates using the color 
criteria described in \S3 from 17,343 sources detected over the 465 
arcmin$^2$ in our survey.  We find an LBG color candidate number
density of 0.79$\pm0.25$ arcmin$^{-2}$ in the five fields with
$u'$BVRI imaging and 1.45$\pm0.66$ for six fields with $u'$VRI
imaging.  

We obtained spectroscopic observations of 529 LBG candidates and 
identified 339 redshifts in which 211 are secure and probable $z>2$ 
LBGs with a distribution of $\langle z\rangle=3.02\pm0.32$.  Of the
211 LBGs, $46\%$ display Lyman $\alpha$ in emission and the remaining 
display Lyman $\alpha$ in absorption.  We find no bias in the 
confidence of our LBG identifications with magnitude and no
bias in magnitude based on individual LBG spectral features. 
Additionally, we find no statistically significant difference in the 
redshift distributions between LBGs identified by spectral features
yet we find an effect in confidence of our probable identifications 
with redshift.  We obtained spectroscopic data for $\sim67\%$ of the 
photometrically selected LBG color candidates in nine fields and 
identified $63\%$ of the obtained spectra.  We find an LBG number 
density of $0.37\pm0.21$ arcmin$^{-2}$ for secure LBGs and
$0.56\pm0.27$ arcmin$^{-2}$ for the combined set of secure and
probable LBGs.

We modeled the photometric selection function based on the redshifted 
galaxy colors from the theoretical galaxy templates used in this 
survey and included photometric uncertainties modeled from the data.  
We combined the effects on the color selection caused by the 
characteristics of the CCDs and $u'$BVRI filters and assumed a uniform 
LBG background redshift distribution.  The modeled data is in
excellent agreement to that of the observed data and is well described 
by a Gaussian function using the parameters of the modeled data 
distribution.  Moreover, a Gaussian function fits the observed
redshift distributions of the data as well using the parameters from 
each of the observed distributions.  We used these results to validate 
the assumption of a uniform or slowly varying LBG background number 
density from $2.4<z<3.6$.  The spatial information of the 211 
spectroscopically identified LBGs from this survey will be used to 
measure the DLA-LBG cross-correlation with the 11 known DLAs in Paper 
II. 
 
Following the success of the Lyman break photometric selection method 
demonstrated by \citet {s96}, several surveys for LBG candidates have 
extended the Lyman break selection method to conventional UBVRI
filters \citep {o01, bl03, f98, r04}.  These surveys rely on photometric 
redshifts to determine galaxy spatial distributions and have a typical
accuracy of $\Delta z\sim0.2$.
The survey presented here provides deep, low resolution spectroscopy 
using the 10 meter Keck telescopes to measure the spatial distribution
of LBGs at $z\sim3$ to $\Delta z\sim0.003$.  The 339 identified
spectra obtained from this survey are used to calibrate the overall 
efficacy of an LBG photometric selection method using conventional 
$u'$BVRI filters. From our spectroscopic sample and conservative
identifications, we find that the criteria for the $u'$BVRI filters 
yield a high LBG detection efficiency (79\%~$z>2$ identified 
redshifts).  The inclusion of objects within 0.2 magnitudes of the
imposed $u'$BVRI color criteria, made to compensate for photometric
uncertainties, allowed for a near doubling of the number of $z>2$
objects with little loss in efficiency (75\%~$z>2$ identified 
redshifts).  Although the  use of all five filters with the color
criteria presented in this paper may be too restrictive in detecting 
the full population of LBGs at $z\sim3$, we find the use of a subset
of these filters ($u'$VRI, $u'$BRI,$u'$BR) allow the inclusion of more 
objects as LBG candidates but are less efficient (69\%, 63\%, and 62\%
~$z>2$ identified redshifts respectively). 

We used the spectroscopic information to offer a possible 
refinement to our original $u'$BVRI color criteria to increase the 
efficiency and provide insight into the entire population of LBGs at 
$z\sim3$.  Although a spectroscopic analysis of the serendipitous 
objects suggests that the original color criteria imposed by our 
selection method detects most of the objects at $z\sim3$, they may not 
include all objects such as galaxies that are highly reddened.  Several 
LBGs selected exclusively from a subset of the $u'$BVRI filters were 
detected outside the regions chosen for the original color cuts.  To 
obtain a more complete census of objects at $z\sim3$, modifications 
to the existing color cuts are suggested.

%% The \notetoeditor{TEXT} command allows the author to communicate
%% information to the copy editor.  This information will appear as a
%% footnote on the printed copy for the manuscript style file.  Nothing 
%% will appear on the printed copy if the preprint or
%% preprint2 style files are used.

%---------------------------------------------------------------------
\acknowledgments

We thank Alan Dressler, Riccardo Giovanelli, and Jeremy Darling
for kindly providing access to the Palomar Observatory and Alan
Dressler and Jeremy Darling for their assistance with the COSMIC
camera and deep image observations of the first two fields in this
survey.  We also thank Chuck Steidel for access to the $U_n$ filter at 
both the Palomar and Keck Observatories that made the important
initial deep images in this bandpass possible and Kurt Adelberger and
Alice Shapley for data regarding the $U_n$ filter transmission. We are 
grateful to the staffs of the Palomar and Keck Observatories for their 
kind assistance with the observations and deft handling of unpredictable 
instrument crises.  We thank the teams behind the $COSMIC$ imager and 
$LRIS$ imaging spectrometer.  Without these instruments, this science 
would not be possible. We also like to thank Judy Cohen and Drew
Phillips for use of their astrometric software designed specifically
for these instruments, Daniel Stern, Andrew Bunker, and Adam Stanford
for the access to their $LRIS$ slitmask reduction routine, and
Hsiao-Wen Chen for graciously allowing the use of a routine that aided 
in the creation of the stacked images.  This work was partially
supported by the National Science Foundation grant AST-0307824 as well 
as the NSF Astronomy \& Astrophysics Postdoctoral Fellowship (AAPF) 
grant AST-0201667 awarded to Eric Gawiser. Finally, the authors wish
to recognize and acknowledge the very significant cultural role and 
reverence that the summit of Mauna Kea has always had within the 
indigenous Hawaiian community.  We are most fortunate to have the 
opportunity to conduct observations from this mountain.

%  we appreciate and
%  hold reverent the mountain of Mauna Kea for its cultural role within 
%  the ancestral Hawaiian community.  We are honored to have had the 
%  opportunity to conduct scientific observations from its sacred summit.

%More information on the AASTeX macros package are available at
%\url{http://www.aas.org/publications/aastex} or the
%\anchor{ftp://www.aas.org/pubs/}{AAS ftp site}.
%For technical support, please write to
%\email{aastex-help@aas.org}.

\clearpage

%---------------------------------------------------------------------------
%   REFERENCES
%---------------------------------------------------------------------------

%---------------------------------------------------------------------------
%   TABLES
%---------------------------------------------------------------------------

\clearpage
\begin{deluxetable}{lcccll}
\tabletypesize{\scriptsize}
\tablecaption{Target QSO Fields with $z\sim3$ DLAs \label{coords}}
\tablewidth{0pt}
\tablehead{
\colhead{Field} & \colhead{R.A.(J2000.0)} & \colhead{Dec.(J2000.0)} &
\colhead{$z_{QSO}$} & \colhead{$z_{DLA(s)}$} & \colhead{log N(HI)}\\ 
\colhead{} & \colhead{$(h~~m~~s)$} & \colhead{$(d~~m~~s)$} 
& \colhead{} & \colhead{} & \colhead{}}
\startdata
LBQS0056+0125 & 00 59 17.62 & +01 42 05.30 & 3.16 & 2.775 
 & $21.0$\tablenotemark{1}\\
%\pm0.1
PKS0336--017  & 03 39 00.65 & -01 33 19.20 & 3.20 & 3.062 
 & $21.2$\tablenotemark{2}\\ 
%\pm0.1
PSS0808+5215  & 08 08 49.43 & +52 15 14.90 & 4.45 & 2.936,~3.113 
 & $20.9$\tablenotemark{3},~$20.7$\tablenotemark{4}\\
%\pm0.3
%\pm0.1
PSS0957+3308  & 09 57 44.50 & +33 08 23.00 & 4.25 & 3.280,~4.180 
 & $20.5$\tablenotemark{4}~, $20.7$\tablenotemark{4}\\
%\pm0.1
%\pm0.2
BRI1013+0035  & 10 15 48.96 & +00 20 19.52 & 4.41 & 3.103 
 & $21.1$\tablenotemark{5}\\
%\pm0.1
PSS1057+4555  & 10 57 56.39 & +45 55 51.97 & 4.12 & 3.050,~3.317 
 & $20.3$\tablenotemark{5}~$^{,}$\tablenotemark{6}~, $20.3$\tablenotemark{7}\\
%\pm0.1
%\pm0.1
PSS1432+3940  & 14 32 24.90 & +39 40 24.00 & 4.28 & 3.272 
 & $21.3$\tablenotemark{4}\\
%\pm0.1
PC1643+4631A  & 16 45 01 09 & +46 26 16.44 & 3.79 & 3.137 
 & 20.7\tablenotemark{8}\\
JVAS2344+3433 & 23 44 51.25 & +34 33 48.64 & 3.01 & 2.908 
 & $21.1$\tablenotemark{4}
%\pm0.1
\tablenotetext{1}{\citet{w95}}
\tablenotetext{2}{\citet{p01}}
\tablenotetext{3}{From this work}
\tablenotetext{4}{\citet{p03}}
\tablenotetext{5}{\citet{slw00}}
\tablenotetext{6}{\citet{celine01}}
\tablenotetext{7}{\citet{lu98}}
\tablenotetext{8}{\citet{sch91}}
\enddata
\end{deluxetable}

\clearpage
%\begin{deluxetable}{lccccc}
%\tabletypesize{\scriptsize}
%\tablecaption{Vega magnitudes using $AB_{95}$ and Johnson-Cousins 
%\label{tbl-2}}
%\tablewidth{0pt}
%\tablehead{
%\colhead{  } & \colhead{$u'$} & \colhead{$B$} & \colhead{$V$} & 
%\colhead{$R_C$} & \colhead{$I_C$}}
%\startdata
%$AB_{95}$ & 0.72  & -0.12  &  0.02  & 0.21  & 0.45 \\
%$J-C$     & 0.02  &  0.03  &  0.03  & 0.03  & 0.02 \\
%\hline
%\hline\\
%$J-C$ to $AB_{95} $  & +0.70  & -0.15  & -0.01  & +0.18  & +0.43 \\
%\enddata
%\end{deluxetable}
%
%\clearpage

\begin{deluxetable}{llccccc}
\tabletypesize{\scriptsize}
\tablecaption{Details of Imaging Observations \label{imaging}}
\tablewidth{0pt}
\tablehead{
\colhead{Field} & \colhead{Instrument and Date\tablenotemark{a}} 
& \colhead{Filter} & \colhead{Exp.(s)} & \colhead{FWHM} 
& \colhead{Depth $AB$\tablenotemark{b}} & \colhead{$E(B-V)$}}
\startdata
LBQS0056+0125 & $LRIS$  Dec 2001            & $U_n$&  9000 & 1.5 & 28.9 & 0.03\\
          &$LRIS/LRISb$ Dec 2001/Feb 2003  & B & 1530/1300 & 2.0/1.2 & 28.1 & \\
              & $LRIS$  Dec 2001/Dec 2002  & V &  3060/600 & 1.4/1.5 & 28.2 & \\
&$LRIS$ Dec 2001/Feb 2003 & R$_s$\tablenotemark{c}~/R& 2800/600 & 1.4/1.7 & 28.3 & \\
              & $LRIS$  Dec 2001                   & I &  2800 & 1.5 & 27.8 & \\
PKS0336--017&$LRISb$ Dec 2002/Nov 2000 &$u'$& 7200/2400& 1.3/0.8 & 29.1 & 0.14\\
              & $LRIS$  Sep 1998                   & B &  1500 & 0.6 & 28.9 & \\
              & $LRIS$  Dec 2002                   & V &  2400 & 1.7 & 28.0 & \\
              & $LRIS$  Sep 1998                   & R &  1200 & 0.6 & 28.9 & \\
              & $LRIS$  Sep 1998                   & I &  1080 & 0.5 & 27.6 & \\
PSS0808+5215  & $LRISb$ Dec 2002             & $u'$&  4500 & 1.2 & 29.3 & 0.04\\
          &$LRIS/LRISb$ Nov 2000/Dec 2002 & B &  1200/1680 & 0.9/0.8 & 29.2 & \\
              & $LRIS$  Dec 2002                   & V &  1740 & 1.2 & 28.8 & \\
              & $LRIS$  Nov 2000/Dec 2002 & R &  1200/2600 & 1.1/1.2 & 28.8 & \\
              & $LRIS$  Dec 2002                   & I &  1000 & 1.1 & 27.3 & \\
PSS0957+3308  & $LRISb$ Dec 2002             & $u'$&  4500 & 1.0 & 29.6 & 0.01\\
              & $LRIS$  Jan 2001                   & B &  3645 & 1.0 & 28.9 & \\
              & $LRIS$  Dec 2002                   & V &  1800 & 1.0 & 29.1 & \\
              & $LRIS$  Jan 2001                   & R &  1860 & 1.0 & 28.6 & \\
              & $LRIS$  Jan 2001/Dec 2002 & I &  1400/1000 & 1.1/1.0 & 27.8 & \\
BRI1013+0035  & $LRISb$ Dec 2002             & $u'$&  4600 & 0.9 & 29.7 & 0.03\\
              & $LRIS$  Nov 2000                   & B &  1200 & 1.1 & 28.5 & \\
              & $LRIS$  Dec 2002                   & V &  1400 & 0.9 & 29.0 & \\
              & $LRIS$  Dec 2002                   & R &  1500 & 0.9 & 28.9 & \\
              & $LRIS$  Dec 2002                   & I &  1000 & 0.9 & 27.7 & \\
PSS1057+4555  & $LRIS$  Apr 2001            & $U_n$&  9000 & 0.7 & 29.2 & 0.01\\
              & $LRIS$  Jan 2001                   & B &  3600 & 0.8 & 28.9 & \\
              & $LRIS$  Apr 2001                   & V &  1800 & 0.7 & 28.6 & \\
&$LRIS$Jan 2001/Apr 2001 &R/R$_s$\tablenotemark{c}& 2320/1680 & 0.8/0.7 & 29.0 & \\
              & $LRIS$  Jan 2001/Apr 2001 & I &  1160/1320 & 0.8/0.7 & 28.1 & \\
PSS1432+3940  &$COSMIC$ Apr 2000            & $U_n$& 19800 & 1.0 & 28.7 & 0.01\\
              &$COSMIC$ Apr 2000                   & B & 10800 & 1.1 & 28.8 & \\
              & $LRIS$  Jan 2001                   & V &  3200 & 0.9 & 28.3 & \\
              &$COSMIC$ Apr 2000                   & R &  9000 & 0.9 & 28.5 & \\
              & $LRIS$  Jan 2001                   & I &  1200 & 0.8 & 27.6 & \\
PC1643+4631A& $LRIS$  Apr 2001              & $U_n$&  7300 & 0.9 & 28.8 & 0.02\\
              & $LRIS$  Apr 2001   & B\tablenotemark{d}&  3720 & 0.9 & 28.5 & \\
              & $LRIS$  Apr 2001                   & V &  1800 & 0.8 & 28.4 & \\
              & $LRIS$  Apr 2001  & R$_s$\tablenotemark{c}&  1740 & 0.7 & 28.5 & \\
              & $LRIS$  Apr 2001                   & I &  1200 & 0.7 & 28.1 & \\
JVAS2344+3433 &$COSMIC$ Sep 2000            & $U_n$& 32400 & 1.4 & 29.2 & 0.08\\
              &$COSMIC$ Sep 2000                   & B &  8400 & 1.2 & 28.8 & \\
              & $LRIS$  Jan 2001                   & V &  2400 & 1.2 & 28.9 & \\
              &$COSMIC$ Sep 2000                   & R &  7980 & 1.0 & 28.7 & \\
              & $LRIS$  Jan 2001                   & I &  1200 & 1.3 & 26.7 & \\
\enddata
\tablenotetext{a}{$LRIS$: The Low Resolution Imaging Spectrometer;
$LRISb$: $LRIS$ with the new blue sensitive CCD chips; $COSMIC$: The 
Carnegie Observatories Spectrograph and Multi-Object Imaging Camera.}
\tablenotetext{b}{1$\sigma$ sky fluctuations per seeing element.} 
\tablenotetext{c}{The R$_s$ filter was used instead of, or in addition
to, the R$_{JC}$ in this field.}
\tablenotetext{d}{Johnson B filter convolved with the transmission of
the 460nm dichroic installed on $LRIS$}
\end{deluxetable}

\clearpage
\begin{deluxetable}{lcccccc}
\tabletypesize{\scriptsize}
\tablecaption{LBG Color Candidates \label{colors}}
\tablewidth{0pt}
\tablehead{
\colhead{Field} & \colhead{Area\tablenotemark{a}}
& \colhead{$u'$BR(I)} & \colhead{$u'$VRI} & \colhead{$u'$BVRI}
& \colhead{$u'$VRI LBG} & \colhead{$u'$BVRI LBG} \\
\colhead{} & \colhead{(arcmin$^2)$} & \colhead{Candidates} & 
\colhead{Candidates} & \colhead{Candidates} & 
\colhead{Density\tablenotemark{b}} & \colhead{Density\tablenotemark{c}}}
\startdata
LBQS0056+0125 & 41.96   & ... & 77  & ... & 1.84$\pm0.21 $\tablenotemark{d}& 
                                                     ...\tablenotemark{d}\\
PKS0336--017  & 40.95     &  87 & ...  & ... & ... & ...\tablenotemark{e}\\
PSS0808+5215  & 45.01    &  61 &  54 & 42 & 1.20$\pm0.16$ & 0.93$\pm0.14$\\
PSS0957+3308  & 39.75    &  47 &  34 & 29 & 0.86$\pm0.15$ & 0.73$\pm0.14$\\
BRI1013+0035  & 37.21    &  45 &  19 & 16 & 0.51$\pm0.12$ & 0.43$\pm0.11$\\
PSS1057+4555  & 41.49    &  68 &  81 & 46 & 1.95$\pm0.22$ & 1.11$\pm0.16$\\
PSS1432+3940&87.16/57.40\tablenotemark{f}&165&79&42&1.38$\pm0.15$&0.73$\pm0.11$\\
PC1643+4631A  & 38.08    & ... & 91  & ... & 2.39$\pm0.25$\tablenotemark{d}& 
                                                     ...\tablenotemark{d}\\
JVAS2344+3433 & 93.14/59.62\tablenotemark{f} & 155 & 51 & 40 & 
              0.86$\pm0.12$\tablenotemark{g}& 0.67$\pm0.11$\tablenotemark{g}\\
\enddata
\tablenotetext{a}{Common area to all filters.} 
\tablenotetext{b}{Number of LBG candidates per arcmin$^2$ using $u'$VRI 
filters for comparison to previous surveys.}
\tablenotetext{c}{Number of LBG candidates per arcmin$^2$ using $u'$BVRI 
filters.}
\tablenotetext{d}{LBG candidates were chosen using $u'$VRI filters.}
\tablenotetext{e}{LBG candidates were chosen using $u'$BRI filters.}
\tablenotetext{f}{Area common to the 3 {\em COSMIC} filters ($u'$,B,R) and 
area common to all 5 filters [$COSMIC$~($u'$,B,R) and $LRIS$~(V,I)], 
respectively.}
\tablenotetext{g}{LBG candidates were chosen using $u'$BR filters.}
\end{deluxetable}

\clearpage
\begin{deluxetable}{llccccc}
\tabletypesize{\scriptsize}
\tablecaption{Details of Spectroscopic Observations \label{spectro}}
\tablewidth{0pt}
\tablehead{
\colhead{Field} & \colhead{Date} & \colhead{Slitmask\tablenotemark{a}}
& \colhead{Area\tablenotemark{b}}
& \colhead{Exp.(s)} & \colhead{N$_{Obj}$\tablenotemark{c}}
& \colhead{FWHM\tablenotemark{d}}}
\startdata
LBQS0056+0125 & Nov 2003  & q0056l15  & 41.3  & 4500  & 37 & 0.8\\
              & Nov 2003  & q0056r15  & 41.3  & 2400  & 37 & 0.9\\
PKS0336--017  & Feb 2003  & 295n0336  & 41.3  & 4500  & 41 & 1.1\\
              & Feb 2004  & 405q0336  & 41.3  & 3600  & 28 & 0.8\\
              & Feb 2004  & 420q0336  & 41.3  & 3600  & 27 & 0.9\\
PSS0808+5215  & Feb 2003  & q0808n15  & 41.3  & 5234  & 27 & 0.9\\
              & Nov 2003  & 122q0808  & 41.3  & 4200  & 30 & 0.9\\
              & Feb 2004  & q0808c15  & 41.3  & 4500  & 25 & 0.8\\
PSS0957+3308  & Feb 2003  & 180p0957  & 41.3  & 4500  & 21 & 0.9\\
              & Nov 2003  & 100q0957  & 41.3  & 3600  & 25 & 1.0\\
              & Feb 2004  & 100r0957  & 41.3  & 4500  & 19 & 0.9\\
BRI1013+0035  & Feb 2003  & q1013n13  & 41.3  & 4500  & 26 & 0.9\\
              & Nov 2003  & q1013r15  & 41.3  & 3600  & 22 & 1.0\\
              & Feb 2004  & q1013r15  & 41.3  & 4500  & 22 & 1.0\\
PSS1057+4555  & Dec 2002  & r1057n10  & 41.3  & 3600  & 36 & 1.0\\
              & Dec 2002  & l1057t10  & 41.3  & 6000  & 33 & 1.1\\
PSS1432+3940  & Feb 2003  & q1432p15  & 41.3  & 4500  & 31 & 1.2\\
PC1643+4631A  & May 2003  & q1643r15  & 41.3  & 4800  & 30 & 0.7\\
              & Feb 2004  & 245q1643  & 41.3  & 3400  & 40 & 0.8\\
JVAS2344+3433 & Nov 2000  & q2342l10  & 31.5  & 7200  & 18 & 0.8\\
              & Nov 2000  & q2342c10  & 31.5  & 7200  & 23 & 0.9\\
              & Oct 2001  & q2342r15\tablenotemark{e} & 31.5 & 7200 & 21 & 1.2\\
              & Dec 2002  & 2342p280  & 41.3  & 7200  & 30 & 1.5\\
\enddata
\tablenotetext{a}{Name of slitmask for reference.}
\tablenotetext{b}{Approximate area covered by the slitmasks in
  arcminute$^{2}$.  Slitmasks overlapped in most fields.}
\tablenotetext{c}{Number of slitlets milled to obtain object spectra.
  Some objects were observed on more than one mask to obtain a higher
  signal-to-noise ratio.}
\tablenotetext{d}{Seeing FWHM taken from alignment stars in arcseconds.}
\tablenotetext{e}{These images were degraded due to an optical oil leak 
onto the collimator of $LRIS$. Objects from this slitmask were not 
included in the sample.}
\end{deluxetable}

\clearpage
\begin{deluxetable}{lcccccccc}
\tabletypesize{\scriptsize}
\tablecaption{Spectroscopic Identifications \label{IDs}}
\tablewidth{0pt}
\tablehead{
\colhead{Field}& \colhead{N$_{Obs}$\tablenotemark{a}} & 
\colhead{Star\tablenotemark{b}} &
\colhead{Low-z$_{gal}$\tablenotemark{c}} &
\colhead{Low-z$_{prob}$\tablenotemark{d}} & 
\colhead{High-z$_{prob}$\tablenotemark{e}} & 
\colhead{LBG\tablenotemark{f}}& 
\colhead{QSO\tablenotemark{g}}&
\colhead{Unidentified\tablenotemark{h}}\\
\colhead{(Category)} & \colhead{} & \colhead{(0)} & \colhead{(1)} & 
\colhead{(2)}& \colhead{(3)} & \colhead{(4)} &\colhead{(5)} & 
\colhead{(6)}}
\startdata
%              tot    s   lz  plz  phz   hz    q    u
LBQS0056+0125 & 73 &  3 & 12 & 26 &  9 & 12 &  1 & 10 \\
PKS0336--017  & 73 &  0 &  0 &  5 &  8 & 14 &  1 & 45 \\
PSS0808+5215  & 59 &  0 &  4 & 11 & 13 & 14 &  1 & 16 \\
PSS0957+3308  & 48 &  4 &  6 &  6 &  5 & 16 &  1 & 10 \\
BRI1013+0035  & 39 &  3 &  0 &  5 &  7 & 16 &  0 &  8 \\
PSS1057+4555  & 63 &  0 &  2 &  2 &  5 & 20 &  1 & 33 \\
PSS1432+3940  & 31 &  0 &  5 &  5 &  5 &  2 &  0 & 14 \\
PC1643+4631A  & 73 &  0 &  3 &  6 & 13 & 29 &  0 & 22 \\
JVAS2344+3433 & 70 &  1 & 15 &  4 &  9 &  9 &  0 & 32 \\
Totals       & 529 & 11 & 47 & 70 & 74 &132 &  5 &190 \\
\enddata
\tablenotetext{a}{Total number of LBG candidates observed spectroscopically}
\tablenotetext{b}{Number of identified Galactic stars}
\tablenotetext{c}{Number of spectroscopically secure $z < 2$ galaxies}
\tablenotetext{d}{Number of objects with probable $z < 2$ identifications}
\tablenotetext{e}{Number of objects with probable $z > 2$ identifications}
\tablenotetext{f}{Number of spectroscopically secure $z > 2$ galaxies}
\tablenotetext{g}{Number of identified QSOs with $z > 2$}
\tablenotetext{h}{Number of unidentified objects}
%\tablenotetext{h}{Not used in the $u'$BVRI selection due to non-standard B
%filter.}
%\tablenotetext{i}{Candidates chosen by $u'$BR selection.}
\end{deluxetable}

\clearpage
\begin{deluxetable}{lcccccccc}
\tabletypesize{\scriptsize}
\tablecaption{Color Selection Criteria Spectroscopic Efficiency 
\label{eff}}
\tablewidth{0pt}
\tablehead{
\colhead{Filters\tablenotemark{a}} & 
\colhead{Star\tablenotemark{b}} &
\colhead{Low-z$_{gal}$\tablenotemark{c}} &
\colhead{Low-z$_{prob}$\tablenotemark{d}} & 
\colhead{High-z$_{prob}$\tablenotemark{e}} & 
\colhead{LBG\tablenotemark{f}} & 
\colhead{QSO\tablenotemark{g}} &
\colhead{Unidentified\tablenotemark{h}} &
\colhead{$z>2$\tablenotemark{i}}}
\startdata
$u'$BVRI                           &  0 &  3 & 10 & 12 & 35 & 1 & 15 & 0.79\\
$u'$BVRI $\pm0.2$\tablenotemark{j} &  5 &  9 & 16 & 24 & 61 & 3 & 41 & 0.75\\
$u'$VRI                            &  1 &  3 &  5 &  8 & 11 & 1 & 15 & 0.69\\
$u'$VRI $\pm0.2$\tablenotemark{j}  &  3 & 18 & 27 & 18 & 37 & 1 & 47 & 0.54\\
$u'$BRI                            &  0 &  4 & 12 & 12 & 14 & 1 & 45 & 0.63\\
$u'$BRI $\pm0.2$\tablenotemark{j}  &  2 &  5 & 21 & 19 & 24 & 1 & 69 & 0.61\\
$u'$BR                             &  0 &  6 &  4 &  8 &  8 & 0 & 14 & 0.62\\
$u'$BR $\pm0.2$\tablenotemark{j}   &  1 & 15 &  4 &  9 &  9 & 0 & 28 & 0.47\\
\enddata
\tablenotetext{a}{Objects meet the criteria of the listed filters only}
\tablenotetext{b}{Number of identified Galactic stars}
\tablenotetext{c}{Number of spectroscopically secure $z < 2$ galaxies}
\tablenotetext{d}{Number of objects with probable $z < 2$ identifications}
\tablenotetext{e}{Number of objects with probable $z > 2$ identifications}
\tablenotetext{f}{Number of spectroscopically secure $z > 2$ galaxies}
\tablenotetext{g}{Number of identified QSOs with $z > 2$}
\tablenotetext{h}{Number of unidentified objects}
\tablenotetext{i}{Fraction of secure and probable objects with $z > 2$}
\tablenotetext{j}{Color criteria extended 0.2 magnitudes to
compensate for photometric uncertainties}
\end{deluxetable}

%---------------------------------------------------------------------------
%   FIGURES
%---------------------------------------------------------------------------

\clearpage

\clearpage

\clearpage

\clearpage

\clearpage

\clearpage

\clearpage

\clearpage

\clearpage

\clearpage

\clearpage

\end{document}